\documentclass[preprint]{aastex}

\shorttitle{Flickering mapping of V2051 Oph}
\shortauthors{R. Baptista \& A. Bortoletto}
\slugcomment{To appear in the Astron. J. (July 2004)}

\begin{document}

\title{Eclipse mapping of the flickering sources \\ 
		in the dwarf nova V2051 Ophiuchi 
\footnote{Based on observations made at Laborat\'orio Nacional de
		Astrof\'{\i}sica/CNPq, Brazil.}}
\author{Raymundo Baptista and Alexandre Bortoletto}
\affil{Departamento de F\'{\i}sica, UFSC, Campus Trindade, 88040-900,
        Florian\'opolis, Brazil}
\email{bap@astro.ufsc.br, alex@astro.ufsc.br}

\begin{abstract}
We report on the eclipse mapping analysis of an ensemble of light 
curves of the dwarf nova V2051 Oph with the aim to study the 
spatial distribution of its steady-light and flickering sources.
The data are combined to derive the orbital dependency of 
the steady-light and the flickering components at two different 
brightness levels, named the 'faint' and 'bright' states.
The differences in brightness are caused by long-term variations in 
the mass transfer rate from the secondary star.
The white dwarf is hardly affected by the long-term changes. Its
flux increases by only 10 per cent from the faint to the bright state,
whereas the disk flux raises by a factor of 2.
Eclipse maps of the steady-light show asymmetric brightness 
distributions with enhanced emission along the ballistic stream
trajectory, in a clear evidence of gas stream overflow.
The comparison between the steady-light maps of the faint and bright
states suggests that the quiescent disk responds to changes in mass 
transfer rate in an homologous way.
The ability to separate the orbital dependency of the low- and 
high-frequency flickering components allowed us to identify the existence 
of two different and independent sources of flickering in V2051~Oph.  
The low-frequency flickering arises mainly in the overflowing gas 
stream and is associated to the mass transfer process.
It maximum emission occurs at the position of closest approach of 
the gas stream to the white dwarf, and its spatial distribution 
changes in response to variations in the mass transfer rate.
High-frequency flickering originates in the accretion disk, showing
a radial distribution similar to that of the steady-light maps and 
no evidence of emission from the hot spot, gas stream or white dwarf.
This disk flickering component has a relative amplitude of about 3 
per cent of the steady disk light, independent of disk radius and 
brightness state.
If the disk flickering is caused by fluctuations in the energy 
dissipation rate induced by magneto-hydrodynamic turbulence, its 
relative amplitude lead to a viscosity parameter $\alpha_{cool}
\simeq 0.1 - 0.2$ at all radii for the quiescent disk.  
This value seems uncomfortably high to be 
accommodated by the disk instability model.

\end{abstract}


\section{Introduction} \label{introduction}

One of the long standing unsolved problems in accretion physics is related 
to the cause of flickering. The term is used to name the intrinsic, 
random brightness fluctuations of 0.01--1 magnitudes on timescales from
seconds to dozens of minutes that are seen in the light curves of T~Tauri
stars (e.g., Herbst \& Shevchenko 1999), mass-exchanging binaries (e.g., 
Augusteijn et~al. 1992; Baptista, Bortoletto \& Harlaftis 2002; 
Sokoloski, Bildsten \& Ho 2001; Bruch 2000 and references therein) 
and AGNs (e.g., Garcia et~al. 1999). Flickering is considered a basic 
signature of accretion (Warner 1995).
Although the study of flickering (a time dependent phenomenon) may yield 
crucial information to the understanding of the viscosity is accretion 
disks, flickering is still the least studied and one of the most poorly 
understood of the accretion phenomena.

A promising step towards a better understanding of flickering consists 
in using indirect imaging techniques such as eclipse mapping (Horne 1985;
Baptista 2001) to probe the surface distribution of the flickering 
sources in Cataclysmic Variables (CVs) [e.g., Welsh \& Wood 1995].
In these close binaries mass is fed to a white dwarf (the primary) by 
a Roche lobe filling companion star (the secondary) via an accretion 
disk (for weakly magnetized white dwarfs) or column (for magnetic 
white dwarfs).  A hot spot is expected to form where the gas stream from 
the donor star hits the outer edge of the accretion disk.

In a mass-exchanging binary, the orbital dependency of the flickering 
amplitude can be obtained by measuring the random brightness variations
caused by flickering in an ensemble of light curves (i.e., the scatter 
curve) as a function of the orbital phase. 
Two different approaches exist for this purpose.

In its original form, the `ensemble' method defines the scatter curve 
as the root-mean-square (rms) deviations of the set of light curves 
with respect to the mean orbital light curve (Horne \& Stiening 1985).
This definition yields good estimates of the flickering amplitude
when the gross features of the light curve are stable in time and all 
the variability is caused by flickering.
In the presence of long term brightness variations (i.e., changes on 
scales longer than the orbital period), the rms curve overestimates 
the flickering amplitude because the scatter produced by flickering
is mixed with that caused by the long term brightness changes (e.g., 
Welsh \& Wood 1995). A critical investigation of this problem
is presented in Bruch (2000).
Bennie, Hilditch \& Horne (1996) devised a way to disentangle the
scatter induced by long term variations from that caused by flickering. 
They associated a `reference' out-of-eclipse flux level 
to each light curve in the data set.
For a given phase bin, a plot of the flux of the data points versus 
the reference flux shows that these quantities are correlated.
The slope of the correlation measures the long term change in the
flux at that phase (a unity slope means that the flux at that phase
varies in an 1:1 proportion with the reference flux) whereas the rms 
with respect to the best linear fit gives an unbiased measurement of 
the flickering component.

In the `single' method one quantifies the flickering in a set of 
light curves by measuring the scatter of each individual light curve 
with respect to a smoothed version of the light curve, and by 
combining the results from all light curves to derive a mean scatter 
curve (Bruch 1996). 
This is equivalent of a data filtering process, yielding the 
high-frequency components of the flickering.
An intrinsic limitation of this method is the difficulty in separating 
the high-frequency flickering from rapid brightness changes, 
p.ex., caused by the eclipse, which may lead to the introduction of 
artifacts around sharp mid-ingress and mid-egress eclipse phases.
A careful choice of the smoothing filter cut-off frequency is needed 
in order to avoid this problem.

The two techniques are complementary.  The `ensemble' method samples
flickering at all frequencies. But because the power spectrum density of 
the flickering can be reasonably well described by a power-law 
$P(f) \propto f^{-\alpha}$, with $\alpha\simeq 1-3$ (Bruch 1992), 
an ensemble curve will normally be dominated by the low-frequency 
flickering components.   On the other hand, the `single' 
method produces curves which sample only the high-frequency flickering,
with the cut-off frequency being defined by the smoothing filter width.
Thus, the combination of both methods opens the possibility not only to 
probe the location of the flickering sources but also to separate
low- and high-frequency components of the flickering.
This may be very useful in the case where different flickering 
mechanisms lead to variability with distinct frequencies (e.g., see 
section~\ref{results}).

V2051~Ophiuchi is a dwarf nova, a sub-class of CVs comprised by 
low-mass transfer binaries showing recurrent outbursts 
(caused by a sudden increase in the mass inflow in the disk) 
in which the brightness increases by factor of 5--100 on timescales 
of weeks to months (Warner 1995). 
V2051~Oph is distinguished among other ultra-short period eclipsing 
dwarf nova (i.e., Z~Cha, OY~Car and HT~Cas) by its 
remarkable flickering activity (amplitude of $\ga 30$ per cent),
which is responsible for a variety of eclipse morphologies and usually 
masks the orbital hump associated to anisotropic emission from the hot 
spot (e.g., Warner \& Cropper 1983; Cook \& Brunt 1983).
Based on their light curves, Warner \& Cropper (1983) concluded that
the flickering in V2051~Oph arises mainly from the inner disk regions,
with only a minor contribution from the hot spot. 
Baptista et~al. (1998a) caught the star in an exceptionally low 
brightness state during 1996 in which the flickering activity was mostly
suppressed. They benefited from the clean view of the eclipses of the 
white dwarf and the faint hot spot at that epoch to derive basic 
parameters of the binary (e.g., masses, radii, orbital inclination).

Bruch (2000) applied the `single' method to a large set of light curves
of V2051~Oph to find that the flickering curve shows an eclipse 
`coincident with the white dwarf eclipse and definitely narrower 
than the disk eclipse' as well as an orbital hump indicative that 
the bright spot contributes to the flickering in this star. 
However, the phase resolution of his scatter curve, 
$\Delta\phi=0.01$~cycles, is not enough to adequately resolve the 
shape of the eclipse.

In this paper we present flickering curves of V2051~Oph of high
phase resolution ($\Delta\phi=0.002$~cycles) obtained with the `ensemble' 
and `single' methods, and we apply eclipse mapping techniques to these 
curves to map the flickering sources in this binary.
The observations are described in section~\ref{observa}. The data analysis 
is reported in section~\ref{analysis}. The results are presented in
section~\ref{results} and discussed in section~\ref{discuss}.
A summary of the conclusions is presented in section~\ref{conclui}.

\section{Observations} \label{observa}

Time-series of high-speed CCD photometry of V2051 Oph in the 
B-band were obtained with an EEV camera ($385 \times 578$ pixels, 
0.58 arcsec pixel$^{-1}$) attached to the 1.6~m telescope of Laborat\'orio 
Nacional de Astrof\'{\i}sica, in southern Brazil, from 1998 to 2002.
The CCD camera is operated in a frame-transfer mode, with a 
negligible dead-time between exposures. It has a GPS board which sets 
its internal clock to UTC time to an accuracy better than 10\,ms.  

The observations are summarized in Table~\ref{log}. 
The fourth column shows the time resolution of the observations in seconds
($\Delta t$), the fifth column lists the eclipse cycle number (E) and 
the last column gives an estimate of the quality of the observations.
The seeing ranged from $1.0\arcsec$ to $2.2\arcsec$.
The data comprise 36 eclipse light curves and only includes runs 
while V2051~Oph was in quiescence.
The star went in outburst on 2000 July 31 and on 2002 August 6.
These observations are not included here and will be presented 
in a separate paper.

All light curves were obtained with the same instrumental set and 
telescope, which ensures a high degree of uniformity to the data set.

Data reduction procedures included bias subtraction, flat-field correction,
cosmic rays removal and aperture photometry extraction.  Time-series were
constructed by computing the magnitude difference between the variable 
and a reference comparison star with scripts based on the aperture 
photometry routines of the APPHOT/IRAF package\footnote{IRAF is 
distributed by National Optical Astronomy Observatories, which is 
operated by the Association of Universities for Research in Astronomy, 
Inc., under contract with the National Science Foundation. }.  
Light curves of other comparison stars in the field were also computed 
in order to check the quality of the night and the internal consistency 
and stability of the photometry over the time span of the observations.

The magnitude and colors of the reference star were tied to the 
Johnsons-Cousins UBVRI systems (Bessell 1990) from observations of this 
star and of blue spectro-photometric standard stars (Stone \& Baldwin 
1983) and standard stars of Graham (1982) made on 4 photometric nights.
We used the relations of Lamla (1981) to transform UBVRI magnitudes to 
flux units.  The B-band flux of the reference star was then used to 
transform the light curves of the variable from magnitude difference to
absolute flux.  We estimate that the absolute photometric accuracy of
these observations is about 10 per cent.  On the other hand, the
analysis of the relative flux of the comparison stars of all observations
indicates that the internal error of the photometry is less than 2 
per cent.   The error in the photometry of 
the variable is derived from the photon count noise and is transformed 
to flux units using the same relation applied to the data.
The individual light curves have typical signal-to-noise ratios of
S/N= 40-50 out-of-eclipse and of S/N= 10-20 at mid-eclipse.

The light curves were phase-folded according to the linear ephemeris 
(Baptista et~al. 2003),
\begin{equation}
T_{mid} = BJDD \;\; 2\,443\,245.977\,52 + 0.062\,427\,8634 \times E \, .
\label{efem}
\end{equation}
Small phase corrections of $-0.0018$ (1998), $-0.0012$ (1999), $+0.0000$
(2000), $+0.0010$ (2001) and $+0.0061$ (2002) cycles were further applied
to the data to make the center of the white dwarf eclipse coincident with
phase zero.  These corrections correspond to the measured O--C mid-eclipse
time residual of the average eclipse light curve of each season with 
respect to the above ephemeris (see Baptista et~al. 2003).

The individual light curves of V2051~Oph are shown superimposed in phase 
in Fig.~\ref{fig1}. 
The upper panel depicts light curves of a comparison star of similar
brightness. The constancy of its flux level over the time span of 
the observations attests that all variations seen in the lower panel are
intrinsic to the variable.
V2051~Oph was brighter in 1999 and 2000 by $\simeq 70$ per cent. 
Hereafter we will refer to the 1999+2000 data as to the `bright state'
and to the 1998+2001+2002 data as to the `faint state'.
Thus, the data set of the faint and bright states comprise 20 and 16 
light curves, respectively.
We note that the data of a given observing season (3-5 days long)
consistently belongs to a same brightness state.  Hence, the brightness
changes under consideration here occur on timescales longer than at 
least a few days.
The comparison of the eclipse shape of both groups of light curves
shows that the main difference is an increase in the brightness of
the hot spot in the bright state in comparison with the faint state, 
indicating that significant long-term changes in the mass transfer 
rate from the secondary star occur on time scales of $\la 1$~yr.
\placefigure{fig1}

The scatter around the mean flux in V2051 Oph is perceptibly larger than
that of the comparison star of similar brightness and is caused by 
flickering.
The scatter is larger close to orbital hump maximum (suggesting that the
bright spot contributes to the flickering) and is smaller during eclipse
(indicating that the flickering sources are occulted at these phases). 
This is in line with the results of Bruch (2000).
Remarkably, the scatter of the light curves of the bright state is 
larger than that of the faint state, suggesting a dependency of
flickering amplitude with brightness level.
This prompted us to perform separate analysis for the data of the two
brightness states.

\section{Data analysis} \label{analysis}

\subsection{Light curve construction} \label{constr}

Scatter curves for the faint and bright states were obtained with the 
`ensemble' and `single' methods as follows.

In order to compute the `single' scatter curve we smoothed each light
curve of a given brightness state with a median filter of length 170~s 
followed by a narrow box car filter (running average) of length 30~s,
and subtracted the smoothed curve from the original data. 
The amount of filtering applied to the light curves is defined by two 
opposite requirements: (i) we wish to preserve the lowest possible 
flickering frequency in the light curve after subtraction of the 
smoothed curve, but (ii) we need to avoid the artifacts that appear in 
the scatter curve around the sharp ingress/egress features of the white 
dwarf if too much filtering is applied.
Smaller filter lengths produce noisy, low amplitude scatter curves
because only the highest frequency components remain in the 
scatter curve after the subtraction of the smoothed curve.
On the other hand, larger filter lengths introduce artificial 
`spikes' in the resulting curve because the smoothed curve is no longer 
able to follow the sharp brightness changes that occur at white dwarf 
ingress and egress phases (e.g., see Fig.3 of Bruch 2000).
The curves of the residuals from all individual light curves were 
combined and divided into a set of phases bins, $\phi$. The mean 
scatter curve $\sigma_{tot}(\phi)$ was then computed from,
\begin{equation}
\sigma_{tot}^2(\phi)= \frac{1}{N(\phi)-1} \sum_{i=1}^{N(\phi)}
\left(f_i - \bar{f}_i \right)^2  \;\;\;,
\label{single}
\end{equation}
where $f_i$ and $\bar{f}_i$ are, respectively, the flux of the $i$th
data point in the original light curve and the corresponding flux 
in the smoothed curve, and $N(\phi)$ is the number of data points 
in each phase bin.

To compute the `ensemble' scatter curve we defined a reference 
out-of-eclipse flux, $f_{ref}$ (the mean flux over all the light curve
excluding the primary eclipse and the orbital hump in the phase range
$-0.3$ to $+0.15$~cycles) for each individual light curve.
The ensemble of light curves of a given brightness state is then 
separated into a set of phase bins, $\phi$. 
A linear fit to the $f_i \times f_{ref}(i)$ diagram of each phase bin
yields an average flux (the flux of the fitted function at the median
reference flux), an angular coefficient (which represents the secular
variability at that phase), and a standard deviation with respect to 
the linear fit, $\sigma_{tot}$.  
By repeating the procedure for all phase bins we obtain the median 
orbital light curve (hereafter referred as the steady-light curve), 
the curve of the secular changes and the scatter curve, $\sigma_{tot}$,
\begin{equation}
\sigma_{tot}^2(\phi)= \frac{1}{N(\phi)-1} \sum_{i=1}^{N(\phi)}
\left\{ f_i - \bar{f}[f_{ref}(i)] \right\}^2  \;\;\;,
\label{ensemble}
\end{equation}
where $\bar{f}[f_{ref}(i)]$ is the value of the fitted linear function
at the reference flux of the $i$th data point. 
The angular coefficient is sensitive to the range of values of the
reference flux and to the scatter (i.e., the amount of flickering) at 
a given phase.  Small $\Delta f_{ref}$ and large scatter contribute to
increase the uncertainty in the angular coefficient.  
This is particularly true for the faint state, for which the range 
of $f_{ref}$ values is quite narrow.
In order to improve the determination of the angular coefficient we 
increased the range of $f_{ref}$ values by combining the data sets of both
brightness states and we computed a single linear fit at each phase bin.
Therefore, the curve of the secular changes is the same for both states 
but the scatter curve is computed separately for each brightness state.

Fig.~\ref{fig2} shows the $f_i \times f_{ref}(i)$ diagrams of the 
faint and bright states at three selected phases.
A linear fit provides a good description of the relation between 
$f_i$ and $f_{ref}(i)$ at all phases.
The scatter around the linear fit is different for each data set and
is generally larger for the bright state.
As discussed in section~\ref{introduction}, we note that using the rms
with respect to the mean flux to compute the scatter of a given data set
overestimates the flickering contribution because no account is made of 
the linear trend in the observed fluxes caused by the long term changes.
\placefigure{fig2}

The `single' and `ensemble' scatter curves have major contributions from 
the photon count statistics (Poisson noise, which is largely dominant in
comparison to the sky scintillation) and from the intrinsic flickering 
activity.  The flickering contribution at phase $\phi$ can then be
obtained from the relation,
\begin{equation}
\sigma_{f}(\phi)= \sqrt{\sigma_{tot}^{2}(\phi) - \sigma_{p}^{2}(\phi)} \;\;\;,
\end{equation}
where
\begin{equation}
\sigma_{p}(\phi)= \frac{1}{N(\phi)} \sum_{i=1}^{N(\phi)} r_i  \;\;\;,
\end{equation}
$r_i$ is the Poisson noise (converted to flux units) of the $i$th 
data point, and $\sigma_{tot}^{2}(\phi)$ is obtained from Eqs.(\ref{single})
and (\ref{ensemble}), respectively for the `single' and `ensemble'
curves.  In both cases, we adopt $\sigma_{f}(\phi) = 0$ when
$\sigma_{p}(\phi) \geq \sigma_{tot}(\phi)$. 

The error in the flickering curve $S_{\sigma_{f}}(\phi)$ is derived 
by error propagation including the error in the standard deviation
$\sigma_{tot}$ (e.g., Lupton 1993),
\begin{equation}
S_{\sigma_{f}}(\phi)= \frac{1}{\sigma_{f}(\phi)}
\sqrt{ \frac{\sigma_{tot}^4(\phi)}{2(N(\phi)-1)} + \sigma_{p}^2(\phi) \,
\mathrm{Var}[\sigma_{p}(\phi)]}
\end{equation}
where
\begin{equation}
\hbox{Var}[\sigma_{p}(\phi)]= 
\left( \frac{1}{N(\phi)} \sum_{i=1}^{N(\phi)} r_i^2 \right) - 
\left( \frac{1}{N(\phi)} \sum_{i=1}^{N(\phi)} r_i \right)^2  \;\;\; .
\end{equation}
The value of $S_{\sigma_{f}}(\phi)$ is meaningless when 
$\sigma_{f}(\phi) = 0$.

The steady-light curves, the curves of the secular change, and the 
`ensemble' and `single' flickering curves for the faint and bright states
are shown in Fig.~\ref{fig3}.  

In order to check the consistency of our
results, we repeated the procedure including different subsets of the 
light curves in the analysis with the 'ensemble' and `single' methods 
of each brightness state.  As expected, the uncertainties in the
flickering curves increase by reducing the number of curves in the
subset. However, the results are not sensitive to the particular
subset of light curves included in the analysis (provided that at least
8-10 light curves are included in each subset) and are the same under
the uncertainties. The computed curves will be discussed in detail in
section~\ref{curves}.

\subsection{Eclipse mapping} \label{mem}

The steady-light, secular changes, and flickering curves were analyzed 
with eclipse mapping techniques (Baptista \& Steiner 1993) to solve 
for a map of the disk surface brightness distribution and for the 
flux of an additional uneclipsed component in each case.
The uneclipsed component accounts for all light that is not contained 
in the eclipse map (i.e., light from the secondary star and/or a 
vertically extended disk wind). 
The reader is referred to Rutten, van Paradijs \& Tinbergen (1992) and 
Baptista, Steiner \& Horne (1996) for a detailed description of and 
tests with the uneclipsed component.

Our eclipse map is a flat cartesian grid of $101 \times 101$ pixels 
centered on the primary star with side $2\; R_{L1}$ (where $R_{L1}$ is 
the distance from the disk center to the inner Lagrangian point).
We adopt $R_{L1}= 0.422\; R_\odot$ (Baptista et~al. 1998a).
The eclipse geometry is defined by the mass ratio $q$ and the inclination
$i$. We adopted the parameters of Baptista et~al. (1998a), $q=0.19$ and
$i=83\degr$, which corresponds to an eclipse phase width of 
$\Delta\phi= 0.0662$. This combination of parameters ensures 
that the white dwarf is at the center of the map.

In the standard eclipse mapping method all variations in the light
curve are interpreted as being caused by the changing occultation of 
the emitting region by the secondary star.
Thus, out-of-eclipse brightness changes (e.g., orbital modulation due to
anisotropic emission from the bright spot) have to be removed before the
light curves can be analyzed.
This is done by fitting a spline function to the phases outside eclipse,
dividing the light curve by the fitted spline, and scaling the result to
the spline function value at phase zero. This procedure removes orbital
modulations with only minor effects on the eclipse shape itself.

For the reconstructions we adopted the default of limited azimuthal 
smearing of Rutten et~al. (1992), which is better 
suited for recovering asymmetric structures than the original default 
of full azimuthal smearing (cf. Baptista, Steiner \& Horne 1996).
The reader is referred to Baptista, Harlaftis \& Steeghs (2000) and 
Baptista (2001) for eclipse mapping simulations which evaluate the 
ability of the eclipse mapping method to reconstruct asymmetric 
structures in eclipse maps.

The statistical uncertainties of the eclipse maps are estimated with a
Monte Carlo procedure (e.g., Rutten et~al. 1992).
For a given input data curve a set of 20 artificial curves is generated, 
in which the data points are independently and randomly varied according 
to a Gaussian distribution with standard deviation equal to the
uncertainty at that point.  The artificial curves are fitted with the 
eclipse mapping algorithm to produce a set of randomized eclipse maps. 
These are combined to produce an average map and a map of the residuals 
with respect to the average, which yields the statistical uncertainty at 
each pixel.  The uncertainties obtained with this procedure are used
to estimate the errors in the derived radial temperature and intensity
distributions.

\section{Results} \label{results}

\subsection{Morphology of the light curves} \label{curves}

The large amplitude flickering usually seen in V2051~Oph makes it hard 
to distinguish the white dwarf and the hot spot eclipses in individual 
light curves.
By averaging many individual light curves we were able of reducing the 
influence of flickering in the steady-light curves, leading to a clean
view of the eclipse shape.

The steady-light curves (Fig.~\ref{fig3}a) show the broad eclipse of a 
faint quiescent disk and the sharp ingress and egress of the white dwarf 
and hot spot eclipses -- which occur at orbital phases in agreement with 
the photometric model of Baptista et~al. (1998a).
The flux of the hot spot is larger at ingress than at egress, 
demonstrating that it is the source of the orbital hump.
We modeled the out of eclipse flux of the steady-light curves as 
the sum of a half-sinusoidal function (i.e., the amplitude of the sine 
term is forced to be zero for $\pi < \phi < 2 \pi$) and a constant 
flux level in order to measure the relative amplitude and phase of 
maximum of the orbital hump.
The steady-light curve of the faint state shows a weak orbital hump
(with an amplitude of 12 per cent of the total flux) centered at phase
$\phi_{max}= -0.16 \pm 0.02$~cycles. 
We estimate a hot spot contribution of $f_{hs}(\mathrm{faint})= 0.35\pm 
0.05$~mJy from its ingress feature.
The orbital hump is more pronounced in the bright state, with a 
relative amplitude of 20 per cent and maximum at phase $\phi_{max}=
-0.13 \pm 0.02$~cycles.  The hot spot flux at ingress in this case is 
$f_{hs}(\mathrm{bright})= 0.65\pm 0.05$~mJy.
\placefigure{fig3}

The measured hot spot flux at ingress is larger than the predicted 
flux of the orbital hump at the same phase.  This is an indication 
that the hot spot has an additional isotropic component.
We subtracted the predicted flux of the orbital hump at hot spot 
ingress from the measured hot spot flux at ingress to find isotropic 
components of $0.25 \pm 0.06$~mJy and $0.27 \pm 0.06$~mJy, respectively 
for the faint and bright states.  Thus, the anisotropic component of 
the hot spot emission increases by a factor of 3 from the faint to 
the bright state, whereas the isotropic component remains essentially 
at the same flux level.

The curve of the secular changes (Fig.~\ref{fig3}b) shows a pronounced 
orbital hump and a V-shaped eclipse with ingress coincident with hot
spot ingress and minimum displaced towards positive phases, indicating
a significant contribution from the hot spot and gas stream to the
long-term brightness variations. 
Remarkably, there is no clear sign of the sharp white dwarf 
ingress/egress features in the eclipse, revealing that the white dwarf 
gives negligible contribution to the long-term brightness changes.
The shape of the curve of the secular changes is identical to that 
of the curve of the difference between the bright and faint states
under the uncertainties.  The latter can be obtained by scaling the
former by a factor of 1/3.  This is a direct consequence of the
`ensemble' assumption that the flux at a given phase scales linearly
with the overall brightness level,
\begin{equation}
f(\phi,x)= f_0(\phi) + x . \left(\frac{\partial f}{\partial f_{ref}}
\right)_{\phi} \; .
\end{equation}
The difference between any two brightness levels 
$\Delta f= f_2 - f_1$ is obtained by multiplying the secular change 
$(\partial f/ \partial f_{ref})_{\phi}$
by a scaling factor $x$ independent of phase. 
In fact, the light curve of the faint and bright states themselves 
can be recovered by adding appropriate proportions $x$ of the curve 
of the secular changes to a common baseline light curve $f_0$.  

The `ensemble' curves (Fig.~\ref{fig3}c) are distinct from the 
corresponding steady-light curves. The eclipse is narrower and 
displaced towards positive phases.
There is no perceptible reduction in flux at the ingress phase of the 
white dwarf, although there is a sharp recovery coincident with the
white dwarf egress in the curve of the bright state.
This suggests that this star is not an important contributor
to the low-frequency flickering, at least in the faint state.
The orbital hump is relatively more prominent than in the steady-light 
curves (relative amplitudes of 70 and 40 per cent, respectively for 
the faint and bright states), indicating a significant (anisotropic)
contribution of the hot spot to the low-frequency flickering.
The absolute amplitude of the hump is larger in the bright state than 
in the faint state, but the relative amplitude is smaller.
This indicates that the contribution of the hot spot to the 
low-frequency flickering is diluted by other sources in the bright
state.
The differences in out-of-eclipse level between the two `ensemble' 
curves reflect the larger amplitude of the flickering in the bright
state.  Together with the difference in eclipse shape, this justifies 
the choice to perform separate flickering analysis for each brightness
state.

The lower amplitude of the `single' curves (Fig.~\ref{fig3}d) in
comparison with the `ensemble' curves is in agreement with the 
expectation that the `single' procedure extracts only the low-amplitude,
high frequency flickering components of the light curve.
The eclipse shapes and the gross features of the `single' curves are 
remarkably different from those of the `ensemble' curves.
The orbital hump is weaker in the faint state (relative amplitude of
30 per cent) and almost inexistent in the bright state (amplitude
of less than 10 per cent).  This suggests that the hot spot contributes
little to the high-frequency flickering.
The eclipses are centered at phase zero but are shallow and wider
than the eclipse of the white dwarf, indicating that the high-frequency
flickering is not localized in the vicinities of the white dwarf but 
arises from a larger region.

Our qualitative analysis reveal that flickering of different frequencies
in V2051~Oph arise from different regions and possibly have different
origins.

\subsection{The contribution of the white dwarf} \label{wd}
 
We separated the contribution of the white dwarf from the steady-light 
curves with a light curve decomposition technique (Wood, Irwin \& Pringle
1985) in order to compute eclipse maps only of the quiescent disk.

Figure\,\ref{fig5} illustrates the procedure with the steady-light 
curve of the bright state.
The contact phases of the white dwarf can be identified as rapid changes 
in the slope of the light curve (Fig.~\ref{fig5}a).
The original light curve is smoothed with a median filter (3 data points)
and its numerical derivative is calculated. The derivative curve is 
smoothed with the same median filter as above to reduce noise and
improve the detection of the white dwarf features.
The ingress/egress of the white dwarf are seen as those intervals where 
the derivative of the light curve is significantly different from zero
(Fig.~\ref{fig5}b).
A spline function is fitted to the remaining regions in the derivative 
to remove the contribution from the extended and slowly varying eclipse 
of the disk. Estimates of the white dwarf flux are obtained by integrating 
the spline-subtracted derivative at ingress and egress. The light curve 
of the white dwarf is then reconstructed by assuming that its flux is 
zero between ingress and egress and that it is constant outside eclipse
(Fig.~\ref{fig5}c).
The separated light curve of the accretion disk (plus hot spot) is 
obtained by subtracting the reconstructed white dwarf light curve from 
the original light curve (Fig.~\ref{fig5}d).
The eclipse maps of the steady-light shown in section~\ref{maps} were 
computed using the white dwarf subtracted steady-light curves.
\placefigure{fig5}

The derived B-band fluxes of the white dwarf in the steady-light curves
of the faint and bright states are, respectively, $0.45 \pm 0.02$~mJy and 
$0.51 \pm 0.02$~mJy.  We applied the same procedure to the B-band 
light curve of Baptista et~al. (1998a) to find a flux of $0.9 \pm 0.1$~mJy.
The B-band average out-of-eclipse flux at that epoch was $1.4 \pm 
0.1$~mJy, while the average out-of-eclipse level of the steady-light 
curves of the faint and bright states are, respectively,
$1.44 \pm 0.03$~mJy and $2.52 \pm 0.06$~mJy.
Hence, the white dwarf flux in our data is lower by a factor of 2 than 
that of the 'low' brightness state of 1996 -- where accretion has 
probably reduced to a minimum and the bare white dwarf could be seen
thru a residual, optically thin disk (Baptista et~al. 1998a).

Because accretion is expected to increase the surface temperature of 
the white dwarf (Godon \& Sion 2003 and references therein), 
it is hard to explain the reduction in flux in terms of a lower 
white dwarf temperature during the epoch of our observations 
(where accretion was certainly taking place) in comparison with the
'low' state of 1996 (where only residual accretion may have occurred).
On the other hand, the discrepancy is plausibly explained if the 
inner disk regions were optically thick at the epoch of our 
observations and veiled the lower white dwarf hemisphere from view, 
thereby leading to a reduction in white dwarf flux by the observed factor.
This is in agreement with the results of Vrielmann, Stiening \& Offutt 
(2002) which found that the accretion disk of V2051~Oph in quiescence 
consists of an optically thin, low density chromospheric layer 
(responsible for the emission lines) on top of a denser and cooler 
opaque disk gas.

Since we have the white dwarf flux at only one wavelength it is not
possible to simultaneously fit its temperature and solid angle.
Therefore, we adopted $R_{wd}= 0.0103 \; R_\odot$ (Baptista et~al. 1998a) 
and a distance of 146~pc (Vrielmann et~al. 2002), and assumed a fixed 
white dwarf solid angle of $\theta^2= 7.82 \times 10^{-3} \; 
[R_{wd}/0.0103\,R_\odot]^2 \; [d/0.146\, Kpc]^{-2} \; (R_\odot/Kpc)^2$
for our observations and twice that value for the observations of 
Baptista et~al. (1998a).
We fitted the observed flux densities from synthetic photometry with
white dwarf atmosphere models (G\"ansicke, Beuermann \& de Martino 1995)
to find $T_{wd}= (24.2 \pm 0.6)\times 10^3$~K and $(25.8 \pm 0.5)\times 
10^3$~K, respectively for the faint and bright states.
The inferred temperature for the `low' brightness state of Baptista et~al. 
(1998a) is the same as that for the faint state, since the measured white 
dwarf flux and the solid angle are scaled by exactly the same factor.

The derived temperatures are significantly larger than the previous
estimates of $\simeq 15000$~K (Catal\'an et~al. 1998; Steeghs et~al. 
2001) and 19600~K (Vrielmann et~al. 2002).  
This seems contradictory because V2051~Oph was 60 per cent brighter at 
the epoch of the observations of Vrielmann et~al. (2002) in comparison
with our observations, whereas the data of Catal\'an et~al. (1998) 
correspond to the 'low' brightness state of Baptista et~al. (1998a)
and those of Steeghs et~al. (2001) are contemporary with our 1998
observations.  
On the other hand, the inferred white dwarf temperatures can be made
compatible with the values derived by those authors if the distance
is reduced to about 100~pc.
Independent estimates of the distance to V2051~Oph would be helpful
to clarify this point.

The difference in white dwarf temperature between the faint and bright 
states is small but significant at the 3-$\sigma$ level. 
It is possibly the result of extra accretional heating in 
consequence of the larger mass transfer rate of the bright state
(e.g., Godon \& Sion 2003).

\subsection{The hot spot and the disk radius}

An estimate of the disk radius can be obtained from the ingress and 
egress phases of the hot spot ($\phi_{hi},\phi_{he}$) under the 
assumption that the hot spot is located where the stream of 
transferred matter hits the outer edge of the accretion disk. 
Measuring hot spot phases in V2051~Oph is not as straightforward as 
measuring white dwarf eclipse phases because the hot spot is less 
compact than the white dwarf (see Fig.~\ref{fig5}).  
We can nevertheless use the derivative technique of section~\ref{wd} 
to estimate the ingress/egress phases of its brightest part. 

The measurements were made on average light curves of each observing 
season (which comprise time intervals of a few days) in order to 
minimize the blur in the ingress/egress phases caused by possible
changes in disk radius with time. The results are consistent for
the data of a given brightness state and are also similar for the
faint and bright states. We find $\phi_{hi}= -0.017\pm 0.002,\; 
\phi_{he}= +0.080 \pm 0.002$ for the faint state and 
$\phi_{hi}= -0.018\pm 0.002 \; , \phi_{he}= +0.082 \pm 0.003$
for the bright state.
The circles that pass thru the points defined by these pairs of 
phases are $R_{hs}= 0.46 \pm 0.02 \; R_{L1}$ and 
$R_{hs}= 0.47 \pm 0.02 \; R_{L1}$, respectively for the faint
and bright states.
This is perceptibly smaller than the disk radius of $0.56 \pm 0.02
\; R_{L1}$ found by Baptista et~al. (1998a) for the `low' brightness
state.

The radial position of the hot spot is similar in the faint and 
bright states, but its flux increased by a factor of about 2 from 
one brightness state to the other (Section~\ref{curves}).
The luminosity of the hot spot is inversely proportional to 
its distance to the disk center and directly proportional to the 
rate at which mass in injected into the outer accretion disk.  
Thus, the only sensible way on changing the flux of a hot spot 
at a fixed radial position is by varying the mass injection 
rate (i.e., the mass transfer rate).
In addition, inspection of the curve of the difference between the 
steady-light curves of the faint and bright states (Fig.~\ref{fig3}a)
confirms that the change in brightness level is mainly the consequence
of enhanced emission from the hot spot and gas stream region in the 
bright state. This underscores the conclusion drawn in 
section~\ref{observa} that the long term brightness changes are 
caused by variations in mass transfer rate.

\subsection{Steady-light and flickering disk structures} \label{maps}

The maps of the surface brightness of the steady-light, secular
changes and flickering are shown in Fig.~\ref{fig6}.  
Circles of radii $0.27\,R_{L1}$ (the circularization radius for 
V2051~Oph, see Baptista et~al. 1998a) and $0.47\,R_{L1}$ (the disk 
radius for the bright state) are depicted in each eclipse map.
Average radial intensity distributions for the eclipse maps in 
Fig.~\ref{fig6} are shown in Fig.~\ref{fig8}. The circularization 
radius ($R_{circ}$) is indicated in the `ensemble' panel.  
Dotted lines in the other panels depict the slope of a power-law 
radial intensity dependency $I(R) \propto R^\alpha$ in each case.
\placefigure{fig6}

The asymmetric steady-light eclipse shapes, with minima displaced
towards positive phases, lead to asymmetric brightness distributions
with enhanced emission along the gas stream region. 
This brightness enhancement traces the ballistic stream trajectory.
It goes well beyond the outer disk rim and can be clearly seen down 
to the region where the stream trajectory passes `behind' the white 
dwarf (i.e., at the disk side farther away from the secondary star).
This is a clear evidence of gas stream overflow in V2051~Oph.

The maximum emission along the stream does not lie on the hot spot
location but at the position of closest approach to the white dwarf.  
This can be explained if the gas stream 
emission is caused by the progressive release of gravitational 
potential energy of the infalling gas (into heat and, thereafter, 
radiation), which has a maximum at the closest distance to the 
central source. An alternative interpretation is that the enhancement
is caused by reprocessing of radiation from the inner disk regions
on a vertically extended bulge along the gas stream trajectory.
In this model, the increased brightness at the region of closest 
approach to the white dwarf could be the consequence of the larger 
solid angle of this region as seen from the illuminating source.

The hot spot and the gas stream region are more pronounced in the 
bright state map, in agreement with the inferred higher inflow
of mass from the companion star in this brightness level.
Aside from the difference in brightness level, the two eclipse maps 
are very similar.  Both show a flat radial intensity distribution in 
the inner disk regions ($I \propto R^{-0.3}$) which decreases 
sharply for $R\ga 0.23\; R_{L1}$ (with a slope $I \propto R^{-2}$).

Not surprisingly, the map of the secular changes is very similar to 
the steady-light maps, with a flat brightness distribution in the inner 
regions and enhanced emission along the gas stream.  In fact, its radial
intensity distribution is hardly distinguishable from that of the
steady-light map of the bright state.  The similarities between the 
steady-light maps and the map of the secular changes indicates that 
the quiescent disk of V2051~Oph responds to changes in mass transfer rate 
in an homologous way (except for the gas stream region, which has
a more pronounced increase in brightness than the rest of the disk).
\placefigure{fig8}

The `ensemble' flickering maps are noticeably different from the 
corresponding maps of the steady-light. 
The narrower eclipse shapes tell us that the center of the flickering 
brightness distributions are displaced towards the disk side farther 
away from the secondary star (for the eclipse to be shorter than the
white dwarf eclipse), whereas the offset towards positive phases 
indicate that the brightness distributions are skewed towards the 
trailing side of the disk (the one containing the gas stream 
trajectory).
The `ensemble' map of the faint state shows an arc-shaped structure 
running along the gas stream trajectory and extending for almost all 
azimuths at a radial position coincident with the circularization 
radius.  There is no evidence of emission from the vicinities of the
white dwarf (the decrease in flickering amplitude at disk center is
statistically significant at the 10-$\sigma$ level, see ahead).
The `ensemble' map of the bright state similarly shows an arc-shaped
structure along the gas stream trajectory, with the brightness 
enhancement starting at a larger radius in comparison with the 
faint state map. In contrast with the faint state map, the map of the 
bright state indicates a clear contribution from the disk center.

In analogy with the steady-light maps, the maximum of the flickering
amplitude along the stream trajectory on both `ensemble' maps does not
occur at the location of the hot spot but around the position of 
closest approach to the white dwarf.
Hence, the conclusion of Warner \& Cropper (1983) that the flickering 
in V2051~Oph arises in general `in the inner disk region with only a 
minor contribution from the hot spot' is consistent with our findings.

The `ensemble' maps show that the low-frequency flickering in V2051~Oph
is associated mostly to the hot spot and gas stream, with additional 
contribution from the innermost disk regions in the bright state.

The `single' maps are remarkably different from the `ensemble' maps,
but bear resemblance with the steady-light maps.
The broad and wide eclipse shapes of the `single' curves map into
extended, shallow and fairly axi-symmetric brightness distributions with 
no clear evidence of the hot spot, gas stream or white dwarf. 
In fact, the `single' maps are slightly asymmetric in the sense that
there is a dearth of emission along the gas stream trajectory --
the high-frequency flickering amplitude is smaller at the regions 
where the low-frequency flickering is stronger.  The `single' radial 
brightness distributions (Fig.~\ref{fig8}) are very similar to the 
radial distributions of the steady-light, without the asymmetries
caused by the gas stream emission.  As with the steady-light, the 
amplitude of the high-frequency flickering also increases in an 
homologous way with the mass transfer rate.  This indicates that
the high-frequency flickering is associated to the accretion disk.

We used Monte Carlo simulations (section~\ref{mem}) for an assessment 
of the statistical significance of the observed structures in the 
flickering maps.  Each flickering map was divided by the map of the
residuals with respect to the mean map to produce a map of the inverse 
of the relative errors, or a signal-to-noise ratio map (e.g., Harlaftis 
et~al. 2004).  The S/N maps are overploted on the corresponding 
flickering maps as contour lines for S/N=10 and 15 in Fig.~\ref{fig12}.
It can be seen that the statistical significance of the structures in all
flickering maps is larger than the 10-$\sigma$ level.  The `ensemble' 
curves have higher S/N and result in maps with a typical significance 
at the 20-$\sigma$ level.
\placefigure{fig12}

Fig.~\ref{fig9} shows the relative amplitude of the low- and 
high-frequency flickering as a function of radius. These curves are
obtained by dividing the radial intensity distribution of each 
flickering map by that of the steady-light map of the corresponding 
brightness state.  The large scatter with respect to the mean 
amplitude in the `ensemble' case for radii $0.2 < R/R_{L1} < 0.5$ is
caused by the strong asymmetries in the corresponding flickering maps.
\placefigure{fig9}

The `single' flickering shows a relative amplitude of about 3 per 
cent, independent of disk radius and brightness state.  The 
amplitude of the high-frequency flickering scales with the brightness 
of the quiescent steady-light disk, thereby confirming the previous 
conclusion that this flickering component is intimately associated 
to the physics of the accretion disk itself.
This lends support to models which attempt to explain flickering 
in terms of physical processes in the accretion disk, such as 
unsteady release of energy at the disk surface because of
convective, turbulent vertical motions (Geertsema \& Achterberg 1992;
Bruch 1992) or  events of magnetic reconnection in the disk 
chromosphere (Kawaguchi et~al. 2000). 
These models predict a power-law behaviour for the power spectrum of 
this disk flickering, with a rise of power towards lower frequencies
(e.g., Geertsema \& Achterberg 1992).

On the other hand, the relative amplitude of the `ensemble' flickering 
varies significantly with radius and brightness level.  Its maximum
occurs in a range of radii around the circularization radius. The
relative amplitude is consistently larger than that of the high-frequency 
flickering and reaches 26 per cent at $R= R_{circ}$ in the faint state.
The maximum of the distribution moves to a larger radius in the
bright state, probably in response to the increased mass transfer rate.
The clear association between the low-frequency flickering and the 
hot spot/gas stream and its sensitivity to the mass transfer rate
indicates that its origin is connected to the mass transfer process. 
Possible explanations include unsteady (or clumpy) mass transfer from 
the secondary star (Warner \& Nather 1971) and turbulence generated by 
the progressive impact of the overflowing gas stream with the disk 
(Shu 1976).

The uneclipsed component of each curve is shown as an horizontal 
dashed line in the left panels of Fig.~\ref{fig6}.  
It is negligible in all cases, except for the `ensemble' curve of 
the bright state. In this case we find an uneclipsed 
flux of $0.017 \pm 0.006$~mJy, corresponding to 8 per cent of the
total flux.  This may indicate the development of a 
vertically-extended clumpy or turbulent wind from the inner disk 
regions in response to the increase in mass transfer rate.  
The bright stripe running from the disk center towards 
the 'back' side of the disk (i.e., the disk hemisphere farther away
from the secondary star) in the corresponding `ensemble' map may be
the projection onto the orbital plane of the base of this disk wind.
If this is correct, the geometry of this structure suggests that
the disk wind is highly collimated.
Support in favor of this scenario comes from the work of Warner \& 
O'Donoghue (1987), who found evidence of an additional asymmetric, 
variable source of light outside of the orbital plane in V2051~Oph.
Alternatively, it may be that part of the incoming unsteady stream 
of gas is flung out of the binary towards the L2 point (to the left of 
the eclipse maps in Fig.~\ref{fig6}) in analogy to the `magnetic 
propeller' scenario of Horne (1999), in which infalling blobs of plasma
are accelerated by the magnetic field of a spinning white dwarf.
The recent discovery of a coherent 52~s oscillation in V2051~Oph
(Steeghs et~al. 2001) suggests that this system may indeed harbor
a magnetic, spinning white dwarf.

\subsection{The flickering power density spectrum}

So far we have concentrated on the analysis of the data in the time
domain.  Here, we investigate the dependency of the 
flickering in V2051~Oph with frequency and we use the results to 
check the consistency of our previous separation of the flickering 
into low- (`ensemble') and high- (`single') frequency components.

The average steady-light curve of the corresponding brightness state
was subtracted from each individual light curve to remove the DC
component and a Lomb-Scargle periodgram (Press et~al. 1992) was 
calculated.  
The periodgrams of all light curves of a given brightness state were
combined to yield a mean periodgram and a standard deviation with 
respect to the mean.  Fig.~\ref{fig10} shows the average power 
density spectrum (PDS) of the faint and bright states binned to a
resolution of 0.02 units in log(frequency).  The dotted lines 
indicate the 1-$\sigma$ limits on the average power in each case.
The PDS show a flat distribution at low frequencies and are well 
described at high frequencies by power-laws $P(f) \propto f^{\alpha}$ 
with $\alpha=-1.2$ and $-1.7$, respectively for the faint and bright 
states.  The frequency below which each distribution becomes flat is 
$f_c(faint)= 7 \times 10^{-4}$~Hz ($t_c= 23.8$~min) and 
$f_c(bright)= 2 \times 10^{-3}$~Hz ($t_c= 8.3$~min).
The slopes of the PDS distributions are consistent with those seen
in other CVs, which can be well described at high frequencies by power 
laws with an average exponent of $\alpha= -2.0 \pm 0.8$ (Bruch 1992).
\placefigure{fig10}

The results of section~\ref{maps} lead us to associate the disk-related 
flickering to the power-law PDS region and the stream-related flickering 
to the flat PDS region.  
The low-frequency cut-off of the filter applied in the `single' procedure
is indicated in Fig.~\ref{fig10}.  The cut-off frequency is such that 
the resulting `single' curves frame only the power-law region of the PDS
(the disk flickering), whereas the `ensemble' curves are dominated by 
the higher power of the regions in which the distribution is flat
(the stream flickering).

We also computed Lomb-Scargle periodgrams of the residual curves
used to derive the 'single' scatter curves (section~\ref{constr}).
These periodgrams were combined to produce average 'single' PDS 
of the faint and bright states.  As expected, the 'single' PDS
show a significant reduction of power at frequencies lower that
the smoothing filter cut-off frequency of the 'single' procedure.
The total, integrated power of the 'single' PDS corresponds to 
30-40 per cent of the integrated power of the corresponding,
unfiltered PDS shown in Fig.~\ref{fig10}, in good agreement with
the flux ratio between the `single' and `ensemble' flickering curves.

\subsection{Radial brightness temperature distributions} \label{trad}

A simple way of testing theoretical disk models is to convert the
intensities in the steady-light eclipse maps to blackbody brightness
temperatures, which can then be compared to the radial run of the effective
temperature predicted by steady state, optically thick disk models.
As pointed out by Baptista et~al. (1998b), a relation between the 
monochromatic brightness temperature and the effective temperature is
non-trivial, and can only be properly obtained by constructing 
self-consistent models of the vertical structure of the disk. 
Since we only have B-band light curves, a detailed disk spectrum 
modeling is beyond the reach with our data set.
Thus, although the brightness temperature should be close to the 
effective temperature for the optically thick disk regions, the
results from this section should be looked with caution.

The disk radial temperature distributions were derived from the 
steady-light maps without the contribution of the white dwarf.
The blackbody brightness temperature that reproduces the observed 
surface brightness at each pixel was calculated assuming a distance of 
146~pc to the system (Vrielmann et~al. 2002). We neglected interstellar 
reddening since there is no sign of the interstellar absorption
feature at 2200~\AA\ in spectra obtained with the Hubble Space
Telescope (Baptista et~al. 1998a).
The disk was then divided in radial bins of $0.05\;R_{L1}$ and a median
brightness temperature was derived for each bin. 
These are shown in Fig.~\ref{fig7} as interconnected symbols.   
The dashed lines show the 1-$\sigma$ limits on the average temperatures.
The larger $\sigma$ values of the bright state distribution reflect
the azimuthal asymmetries in the intermediate disk regions ($0.2-0.3 \; 
R_{L1}$) caused by the enhanced emission along the gas stream trajectory.
Steady-state disk models for mass accretion rates of $\log$ \.{M}$= -9,
-10$, and $-11 \;M_\odot\;$yr$^{-1}$ are plotted as dotted lines for 
comparison.
\placefigure{fig7}

The distributions are flatter than the $T \propto R^{-3/4}$ law for
optically-thick steady-state disks for $R \la 0.23\; R_{L1}$.  
The temperatures in the faint state range from 8000~K in the inner disk
($R= 0.1\;R_{L1}$) to about 4700~K in the outer disk regions 
($R \simeq 0.5\;R_{L1}$).  The temperatures in the bright state are 
higher than those of the faint state by about 1000~K at all disk radii.
The inferred mass accretion rates (in $M_\odot\,$yr$^{-1}$) are 
$\log \dot{\mathrm{M}}_{0.1}= (-11.0 \pm 0.2)$ at $R= 0.1\; R_{\rm L1}$ and 
$\log \dot{\mathrm{M}}_{0.3}= (-10.05 \pm 0.05)$ at $R= 0.3\; R_{\rm L1}$ 
for the faint state, and  $\log \dot{\mathrm{M}}_{0.1}= (-10.8 \pm 0.2)$ 
and $\log \dot{\mathrm{M}}_{0.3}= (-9.92 \pm 0.02)$ for the bright state.
The derived disk temperatures and mass accretion rates are consistent with
(although slightly smaller than) those inferred by Vrielmann et~al. (2002).
We note that the inferred brightness temperatures of the quiescent 
disk of V2051~Oph are systematically higher than those derived for the
dwarf nova OY Car (Wood et~al. 1989) and Z Cha (Wood et~al. 1986) in 
quiescence.
If the disk of V2051~Oph is optically thick, the large range of \.{M} 
required by the brightness temperature distributions implies that gas 
is pilling up in the outer disk regions during quiescence at a rate 
$\simeq 10$ times higher than that at which the gas in the inner disk 
falls onto the white dwarf.

If the distance is reduced to 100~pc, the disk brightness temperatures
become lower by $\simeq 1000$~K in the inner disk regions and by $\simeq
500$~K in the outer disk.  The inferred mass accretion rates will 
become 40 per cent smaller than the above values.

According to the disk instability model, dwarf novae outbursts are
driven by a thermal instability within their disks (e.g., Lasota 2001
and references therein).  In this model, there is a critical effective
temperature, $T_{\rm eff}(crit)$, below which the disk gas should remain 
while in quiescence in order to allow the thermal instability to set in
(e.g., Warner 1995),
\begin{equation}
T_{\rm eff}(crit) = 7432\;\left( \frac{R}{R_{L1}} \right)^{-0.105}
\left( \frac{M_1}{0.78\,M_\odot} \right)^{-0.15} \; K \;\; .
\label{tcrit}
\end{equation}
This relation is plotted in Fig.~\ref{fig7} as a dot-dashed line.
It can be seen that the inferred brightness temperatures of the
quiescent disk of V2051~oph are everywhere below $T_{\rm eff}(crit)$, 
in agreement with the prediction of the disk instability model.
The agreement is better if the distance is reduced to 100~pc.

\section{Discussion} \label{discuss}

\subsection{A comparison with previous results}

The eclipse in the flickering (`single') curve of Bruch (2000) is 
narrower and deeper than that of our `single' curves and led him to 
conclude that the flickering in V2051~Oph originates in the innermost 
disk regions.  This is in contrast with our results.  
Here we explore a possible explanation for this difference.

Bruch's (2000) smoothing filter cut-off frequency of 0.018~Hz implies 
that only flickering occurring on time scales shorter than 1 minute are 
included in his `single' curve, whereas our `single' curves comprise
flickering of time scales shorter than 2.8 min.  Taking into account 
the power-law dependency of the flickering distribution, this means that
his flickering curve samples typically higher frequencies than our
'single' curves.  
Both the disk turbulence and the magnetic reconnection flickering
models predict a spatial segregation of flickering events, with 
higher frequency flickering occurring closer to disk center 
(Geertsema \& Achterberg 1992; Kawaguchi et~al. 2000).  Thus, a
`single' flickering curve sampling higher frequencies should lead
to a narrower and deeper eclipse than a curve sampling lower frequencies
as the distribution of the flickering sources in the former case
is more concentrated towards the disk center.
This could explain the observed difference in shape of
`single' curves with different cut-off frequencies.

We tested this idea by creating 'single' curves using the same cut-off 
frequency of Bruch (2000).  Although the resulting curves are too 
noisy to allow drawing any firm conclusion, we found marginal evidence 
that the increase in cut-off frequency indeed lead to a narrower and 
deeper eclipse shape.  A larger data set (to increase the S/N
ratio of the `single' curves) is required in order to provide a more
conclusive test of this prediction.
 
It is worth mentioning that V2051~Oph was perceptibly brighter at 
the epoch sampled by Bruch's (2000) data (out-of-eclipse magnitude of 
$B\simeq 15.0$~mag) than at the epoch of our observations 
($B\simeq 15.7-16.2$~mag).  We remark that if the flickering 
distribution (and the relative importance of each of the flickering 
sources) changes significantly with brightness level, the comparison 
made in this section may be misleading.
In addition, we note that the flickering curve of Bruch (2000) was
derived from a data set containing a mix of white light observations 
obtained with photomultipliers of different effective wavelengths
(e.g., the `blue' Amperex 56\,DVP vs. the `red' RCA 31034A) 
including both data in quiescence and in outburst (Warner \& Cropper 1983;
Warner \& O'Donoghue 1987) and, therefore, should be looked with some
caution.

\subsection{Estimating the viscosity parameter $\alpha$}

Geertsema \& Achterberg (1992) investigated the effects of 
magneto-hydrodynamic (MHD) turbulence in an accretion disk. They found 
that the rate of energy dissipated per unit area at the disk surface 
shows large fluctuations which could be a source of flickering in CVs
and x-ray binaries.  Here we use their model and the observed relative
amplitude of the `single' flickering to estimate the
viscosity parameter $\alpha$ (Shakura \& Sunyaev 1973) of the
quiescent accretion disk in V2051~Oph.

In the model of Geertsema \& Achterberg (1992), the rms value of the 
fluctuations $\sigma(D)$ in the average rate of energy dissipated per
unit area $\langle D \rangle$ is given by,
\begin{equation}
\frac{\sigma(D)}{\langle D \rangle} \simeq \frac{2.5}{\sqrt{N}} \;\;\; ,
\end{equation}
where,
\begin{equation}
D(r) = \frac{3 G M_1 \hbox{\.{M}}}{8 \pi r^3} 
\left( 1 - \sqrt{\frac{r_1}{r}} \right)  \;\;\; ,
\end{equation}
$M_1$ and $r_1$ are, respectively, the mass and radius of the white dwarf,
and $N$, the number of turbulent eddies that contribute to the local
fluctuation, is given by,
\begin{equation}
N(r) = 4 \pi \frac{r}{H} \left( \frac{H}{L} \right)^2 \simeq 
   1.25 \times 10^3 \left( \frac{H}{L} \right)^2   \;\;\; ,
\end{equation}
where $H$ is the disk scale height, $L$ is the size of the largest 
turbulent eddies, and we assumed $H/r \simeq 10^{-2}$, typical
of thin disks (e.g., Frank, King \& Raine 1992).
The $\alpha$-parameter of Shakura \& Sunyaev (1973) can be written as,
\begin{equation}
\alpha \simeq 0.9 \left( \frac{L}{H} \right)^2  \simeq 
1130 \; \frac{1}{N} \simeq 180 \left[ \frac{\sigma(D)}
{\langle D \rangle} \right]^2 \;\; .
\label{eq.flick}
\end{equation}

If the disk flickering is caused by fluctuations in the energy 
dissipation rate induced by MHD turbulence, the relative amplitude of 
this flickering component is a direct measurement of the ratio 
$\sigma(D)\,/\langle D \rangle$.  
Replacing the observed relative amplitude of the `single' flickering 
in the above expressions we find $N\simeq 7000$, 
$L/H \simeq 0.42 \; (10^2\, H/r)^{-1/2}$, and a viscosity parameter
$\alpha \simeq 0.16 \; (10^2\, H/r)^{-1}$.
Because the relative amplitude of the 'single' flickering is constant
with radius, the inferred viscosity parameter is the same at all disk
radii.

The inferred viscosity parameter is significantly higher than expected
for a quiescent, cool disk in the disk instability model 
($\alpha_{cool} \simeq 10^{-2}$).  In fact, it is of the order of the 
expected viscosity of hot disks during dwarf novae outbursts 
($\alpha_{hot} \simeq 0.1-0.2$, see e.g., Lasota 2001).
Thus, if our assumption that MHD turbulence is the cause of the disk
flickering is correct, the estimated viscosity parameter is 
uncomfortably high to be accommodated by the disk instability model.

One way to avoid this inconsistency is if the accretion disk of 
V2051~Oph is geometrically thick ($H/r \ga 0.05$).  
However, there is a problem with this thick disk scenario.  
It would require a disk half-opening angle of $\beta \ga 3\degr$.  
At the high inclination of V2051~Oph, 
this would lead to an apparent brightness contrast between the emission 
from the disk hemisphere closest to (the `front' side) and farthest from
(the `back' side) the secondary star, with the `back' side being a factor
of at least 2.5 brighter than the `front' side
\footnote{We note that this exercise does not take into account possible
limb darkening effects, which would make the brightness ratio ever
larger than assumed.}.
There is no evidence of front-back brightness contrast in our 
steady-light eclipse maps.

An important implication of the derived high value of the viscosity 
parameter $\alpha$ is that the quiescent disc of V2051~Oph should be in 
a steady-state (because the resulting viscous timescale, $\simeq 30$ days, 
is short in comparison with the typical timescale between outbursts, 
$\simeq 10^{3}$ days).  This is apparently in contradition with the 
flat radial temperature profile derived in section~\ref{trad}. 
However, the optical spectrum of V2051~Oph is dominated by strong Balmer
lines and a Balmer jump in emission, indicating important contribution 
from optically thin gas.
In this case, the assumption that the B-band blackbody brightness
temperature reflects the gas effective temperature probably fails and 
the comparison of the radial temperature distribution with theoretical 
disk models (Fig.~\ref{fig7}) is possibly meaningless.

\section{Summary} \label{conclui}

We applied the `single' and `ensemble' methods to a uniform set of 
light curves of V2051~Oph to derive the orbital dependency of its 
steady-light, long-term brightness changes, low- and high-frequency 
flickering components.
The data can be grouped in two different brightness levels,
named the 'faint' and 'bright' states. 
The differences in brightness level are caused by variations in mass 
transfer rate from the secondary star (by a factor of 2) occurring 
on time scales longer than a few days and shorter than 1~yr.
The white dwarf is hardly affected by these long-term changes. Its
flux increases by only 10 per cent from the faint to the bright state,
whereas the disk flux raises by a factor of 2.

Eclipse maps of the steady-light show asymmetric brightness 
distributions with enhanced emission along the ballistic stream
trajectory, in a clear evidence of gas stream overflow.
The maximum emission along the stream occurs at the position of 
closest approach to the white dwarf.  
The hot spot and the gas stream region are more pronounced in the 
bright state map, in agreement with the inferred higher mass transfer
rate of this brightness level.
Aside of this, the eclipse maps of the
faint and bright states and of the secular changes are similar,
showing a flat radial intensity distribution in the inner disk 
regions ($I \propto R^{-0.3}$) which decreases sharply for 
$R\ga 0.23\, R_{L1}$ (with a slope $I \propto R^{-2}$).
These similarities lead us to conclude that the quiescent disk of 
V2051~Oph responds to changes in mass transfer rate in an homologous way.

Our flickering mapping analysis reveal the existence of two different
sources of flickering in V2051~Oph, which lead to variability at
distinct frequencies.  The low-frequency flickering arises mainly in 
the overflowing gas stream and is connected to the mass transfer 
process. Its maximum emission occurs around the position at which the
stream is closer to the white dwarf and the amplitude reaches $10-25$ 
per cent of the steady-light intensity at the same location.
In the bright state there is additional contribution from the disk 
center and an extra, uneclipsed component of 8 per cent of the total 
flux which may indicate the development of a vertically-extended clumpy
or turbulent wind from the inner disk regions.
Unsteady mass transfer or turbulence generated after the shock between
the stream and the disk material may be responsible for this 
stream flickering component.

The high-frequency flickering is produced in the accretion disk.
It is spread over the disk surface with a radial distribution similar 
to that of the steady-light maps and it shows no evidence of emission 
from the hot spot, gas stream or white dwarf.  This disk flickering 
component has a relative amplitude of about 3 per cent of the 
steady-light, independent of disk radius and brightness state. 
If the disk flickering is caused by fluctuations in the energy 
dissipation rate induced by MHD turbulence, its relative amplitude 
lead to a viscosity parameter $\alpha \simeq 0.1 - 0.2$ at all 
radii for the quiescent disk of V2051~Oph.  
This value seems uncomfortably high to be 
accommodated by the disk instability model.

Previous studies suggested that flickering may originate from two
separate sources, the hot spot and a turbulent inner disk region 
(Warner 1995; Bruch 2000). 
Our flickering mapping experiment reveals a more complex situation,
in which flickering arising
from the inner disk regions may in fact have the same origin as 
the hot spot flickering, namely, the mass transfer process.
Stream flickering from the inner disk regions may be largely 
dominant over the disk flickering if gas stream overflow occurs 
in a quiescent dwarf novae, because the contribution from the 
disk flickering in this case is small (since the disk itself is 
relatively faint).  Also, although the hot spot contributes to 
the flickering, it is a minor source in comparison to the
inner gas stream in this situation.  This seems to be the case
in V2051~Oph.
On the other hand, if the disk flickering amplitude is a fixed 
fraction of the steady disk light, it might become the dominant
flickering source and may appear as arising mainly from the innermost 
disk regions in a high-\.{M}, nova-like system as the surface 
brightness of the bright, opaque disks of these systems 
decreases sharply with radius.

Finally, we note that the measurement of the disk flickering component 
yields a novel and independent way to estimate the disk viscosity 
parameter $\alpha$. 
The large value found in this first experiment seems to favor the 
mass-transfer instability model (Warner 1995 and references therein).

\acknowledgments

This work was partially supported by the PRONEX/Brazil program through
the research grant FAURGS/CNPq 66.2088/1997-2.
RB and AB acknowledge financial support from CNPq/Brazil, respectively 
through grants no. 300\,354/96-7 and 577\,266/1997-7.

\clearpage

\begin{deluxetable}{lccccccc}
  \tabletypesize{\footnotesize}
  \tablecaption{Journal of the observations \label{log}}
  \tablewidth{0cm}
  \tablecolumns{8}
  \tablehead{ \colhead{date} & \colhead{Start} & \colhead{Number of} &
  \colhead{$\Delta t$} & \colhead{E} & \colhead{Phase range} &
  \colhead{brightness} & \colhead{Quality \tablenotemark{a}} \\ [-0.5ex]
  & \colhead{(UT)} & \colhead{exposures} & \colhead{(s)} & \colhead{(cycle)} 
  & & \colhead{state} }

\startdata
1998 Jul 25 & 22:15 & ~283 & 3 & 124535 & $+0.00,+0.31$ & faint & B \\ [-0.5ex]
			& 23:15 & ~888 & 5 & 124536 & $-0.32,+0.50$ & faint & B \\
1998 Jul 26 & 00:29 & ~804 & 5 & 124537 & $-0.50,+0.27$ & faint & B \\ [-0.5ex]
			& 21:40 & ~828 & 5 & 124551 & $-0.37,+0.50$ & faint & A \\ [-0.5ex]
			& 22:58 & 1077 & 5 & 124552 & $-0.37,+0.50$ & faint & A \\
1998 Jul 27 & 00:28 & 1079 & 5 & 124553 & $-0.37,+0.50$ & faint & A \\ [-0.5ex]
			& 01:58 & 1077 & 5 & 124554 & $-0.37,+0.50$ & faint & A \\ [-0.5ex]
			& 03:28 & ~785 & 5 & 124555 & $-0.50,+0.39$ & faint & A \\ [1ex]

1999 Jul 12 & 22:42 & 1098 & 5 & 130174 & $-0.19,+0.83$ & bright & A \\
1999 Jul 13 & 03:02 & ~669 & 5 & 130177 & $-0.28,+0.34$ & bright & A \\
1999 Jul 15 & 21:32 & ~344 & 5 & 130221 & $+0.10,+0.47$ & bright & A \\ [-0.5ex]
			& 22:11 & ~548 & 5 & 130222 & $-0.47,+0.50$ & bright & B \\ [-0.5ex]
			& 23:38 & ~733 & 5 & 130223 & $-0.50,+0.50$ & bright & C \\
1999 Jul 16 & 01:08 & ~114 & 5 & 130224 & $-0.50,-0.34$ & bright & C \\ [1ex]

2000 Jul 28 & 22:37 & ~726 & 5 & 136293 & $-0.19,+0.50$ & bright & A \\ [-0.5ex]
            & 23:39 & ~941 & 5 & 136294 & $-0.50,+0.50$ & bright & A \\
2000 Jul 29 & 01:08 & 1074 & 5 & 136295 & $-0.50,+0.50$ & bright & A \\ [-0.5ex]
          & 02:38 & ~538 & 5 & (136296) & $-0.50,+0.23$ & bright & A \\ [-0.5ex]
            & 21:23 & ~281 & 5 & 136308 & $+0.01,+0.50$ & bright & A \\ [-0.5ex]
            & 22:07 & 1079 & 5 & 136309 & $-0.50,+0.50$ & bright & A \\ [-0.5ex]
            & 23:37 & 1078 & 5 & 136310 & $-0.50,+0.50$ & bright & A \\
2000 Jul 30 & 01:07 & ~860 & 5 & 136311 & $-0.50,+0.30$ & bright & A \\ [-0.5ex]
            & 02:43 & ~848 & 5 & 136312 & $-0.43,+0.36$ & bright & A \\ [-0.5ex]
            & 21:54 & ~986 & 5 & 136325 & $-0.62,+0.29$ & bright & B \\ [1ex]

2001 Jun 25 & 22:34 & ~587 & 5 & 141611 & $-0.05,+0.50$ & faint & A \\ [-0.5ex]
            & 23:24 & ~960 & 5 & 141612 & $-0.50,+0.50$ & faint & A \\
2001 Jun 26 & 00:53 & ~614 & 5/10 & 141613 & $-0.50,+0.64$ & faint & A \\
[-0.5ex]
            & 04:12 & ~359 & 10 & 141615 & $-0.29,+0.38$ & faint & C \\
2001 Jun 27 & 21:38 & ~539 & 10 & 141643 & $-0.64,+0.36$ & faint & B \\ [-0.5ex]
            & 23:23 & ~541 & 10 & 141644 & $-0.47,+0.53$ & faint & B \\
2001 Jun 28 & 01:20 & ~520 & 10 & 141645 & $-0.16,+0.80$ & faint & B \\ [-0.5ex]
            & 05:37 & ~327 & 10 & 141648 & $-0.31,+0.29$ & faint & B \\ [-0.5ex]
            & 23:09 & ~393 & 15 & 141660 & $-0.60,+0.50$ & faint & C \\
2001 Jun 29 & 00:49 & ~303 & 15 & 141661 & $-0.50,+0.34$ & faint & C \\ [1ex]

2002 Aug 04 & 22:28 & ~626 & 10 & 148099 & $-0.66,+0.50$ & faint & B \\
2002 Aug 05 & 00:13 & ~672 & 10 & 148100 & $-0.50,+0.75$ & faint & B \\
\enddata
\tablenotetext{a}{ Sky conditions: A= photometric (main comparison stable), 
B= good (some sky variations), C= poor (large variations and/or clouds) }

\end{deluxetable}
%

\clearpage

%
\begin{figure}	
\caption{ Bottom: light curves of V2051 Oph in 1998, 2001 and 2002 (gray 
   dots) and in 1999 and 2000 (solid symbols). Top: The light curves of a
   comparison star of similar brightness.  Vertical dashed lines mark the
   ingress/egress phases of the white dwarf, whereas dotted lines indicate 
   the ingress phase of the bright spot and the beginning/end of the eclipse 
   of an accretion disk of radius $0.6\,R_{L1}$. The scatter around the mean 
   flux yields an indication of the flickering amplitude at each phase. }
\label{fig1}
\end{figure}
%
%
\begin{figure}	
\includegraphics[scale=0.7]{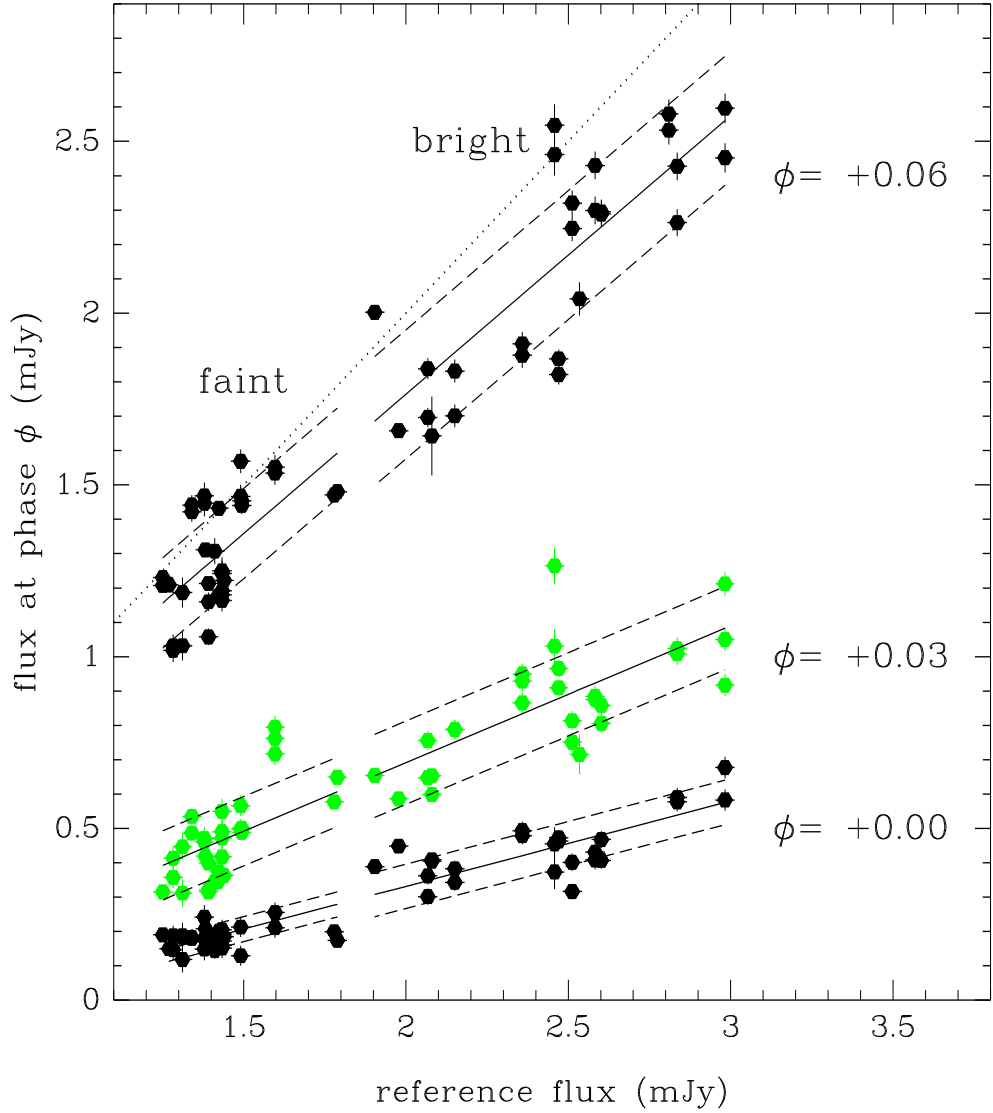}
\caption{ $f_i \times f_{ref}(i)$ diagrams at three selected orbital 
   phases for the faint and bright states (filled symbols with error bars). 
   Solid lines show the best linear fit to the data. Dashed lines indicate 
   the standard deviation (flickering + photon count noise) with respect
   to the fit in each case. A dotted line indicates the unity slope. }
\label{fig2}
\end{figure}
%
%
\begin{figure}	
\caption{ (a) Median (steady-light) curves for the faint (left) and bright
   (right) states. The lower curve in the right panel is the difference
   between the two curves (in the sense bright minus faint).
   (b) The curve of the secular changes, repeated in the right panel for
   visualization purposes. (c) The `ensemble' curves.
   The lower curve in each panel indicates the contribution of the
   Poisson noise, $\sigma_{p}(\phi)$, to the total scatter. (d) The 
   `single' curves; the contribution of the Poisson noise is the same as 
   in (c). The ingress/egress phases of the white dwarf and bright spot
   are indicated, respectively, by dashed and dotted vertical lines. 
   Representative error bars are shown in all panels. }
\label{fig3}
\end{figure}
%
%
\begin{figure}	
\includegraphics[angle=-90,scale=0.6]{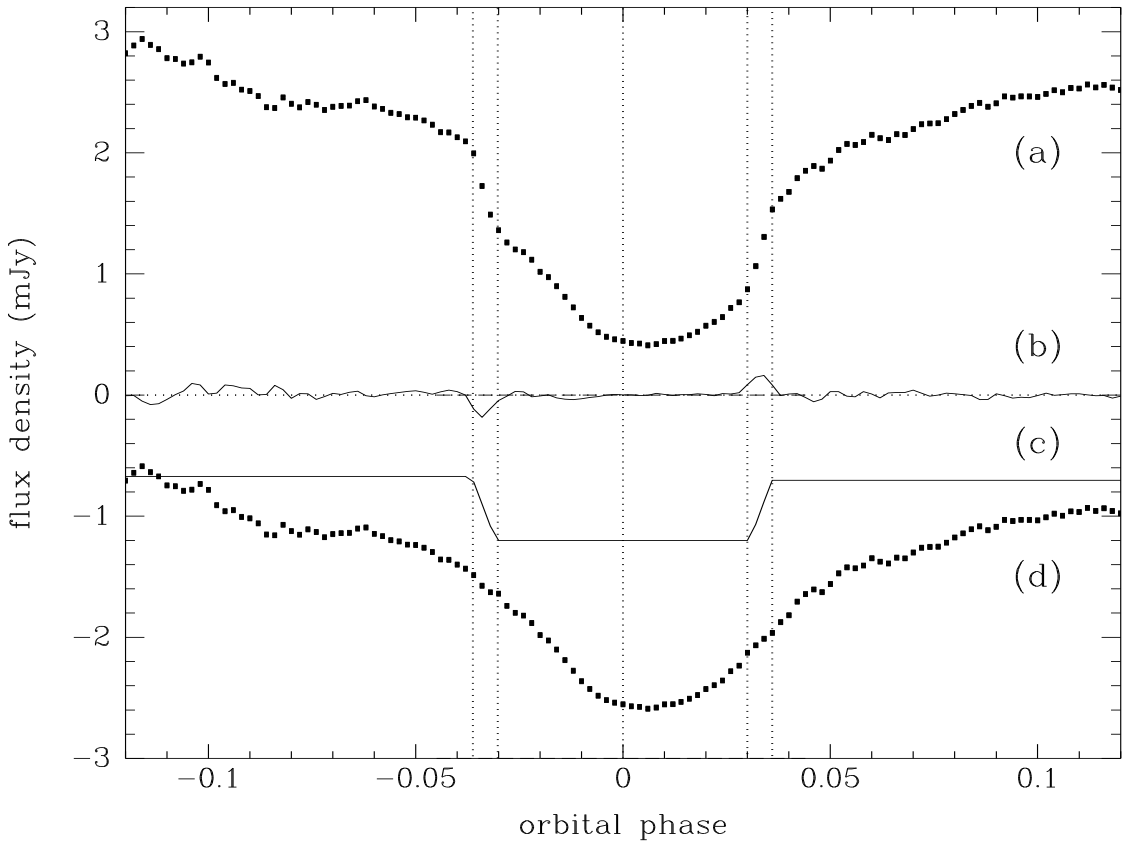}
\caption{ Removing the contribution of the white dwarf from the steady-light
    curve of the bright state. (a) the original light curve; (b) the
    median-filtered derivative of light curve (a) after removing the
    slowly-varying disk component; (c) the reconstructed white dwarf
    light curve, shifted downwards by 1.2~mJy; (d) the light curve 
    without the white dwarf component, shifted downwards by 3~mJy.
    Vertical dotted lines mark the mid-eclipse and the contact phases 
    of the white dwarf. }
\label{fig5}
\end{figure}
%
%
\begin{figure}	
\caption{ Left: the steady-light, secular changes and flickering curves 
  (dots) and the corresponding eclipse mapping models (solid lines). 
  Vertical dotted lines mark the ingress/egress phases of the white 
  dwarf and mid-eclipse.  The curves of the bright state were vertically
  displaced for visualization purposes. 
  Horizontal dotted lines indicate the true zero level and horizontal 
  dashed lines mark the uneclipsed component in each case.
  Right panels: the corresponding eclipse maps in a logarithmic greyscale. 
  Dotted lines depict the primary Roche lobe, the gas stream trajectory,
  and circles of radius 0.27 and $0.47\,R_{L1}$.  The log of intensity 
  scale of each case is indicated in the corresponding greyscale bar. }
\label{fig6}
\end{figure}
%
%
\begin{figure}	
\includegraphics[bb=0cm 1cm 21cm 18cm,angle=-90,scale=0.65]{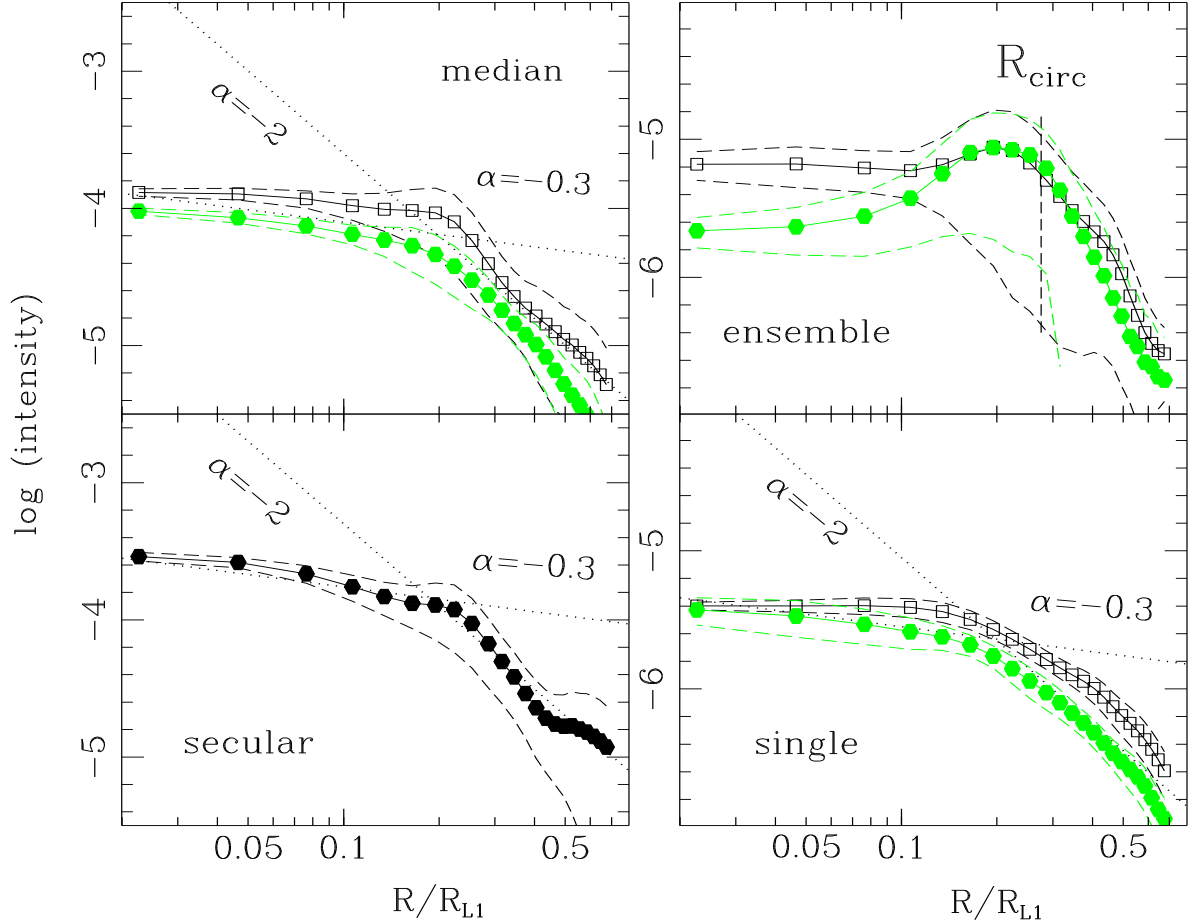}
\caption{ Radial intensity distribution of the steady-light (top 
   left-hand panel), the secular changes (bottom left-hand panel), the 
   `ensemble' (top right-hand panel) and `single' (bottom left-hand panel)
   eclipse maps.  The distributions of the faint and bright states are
   indicated, respectively, by filled circles and open squares.  Dashed 
   lines show the 1-$\sigma$ limit on the average intensity for a given 
   radius.  Abscissae are in units of the distance from disk center to 
   the inner Lagrangian	point $R_{L1}$.  The circularization radius 
   $R_{circ}= 0.27\;R_{L1}$ (Baptista et~al. 1998a) is indicated in the
   `ensemble' panel.  Dotted lines in the other panels depict the slope
   of a power-law radial intensity dependency $I(R) \propto R^\alpha$
   in each case. }
\label{fig8}
\end{figure}
%
%
\begin{figure}	
\includegraphics[bb=0cm 2cm 22cm 18cm,angle=-90,scale=0.75]{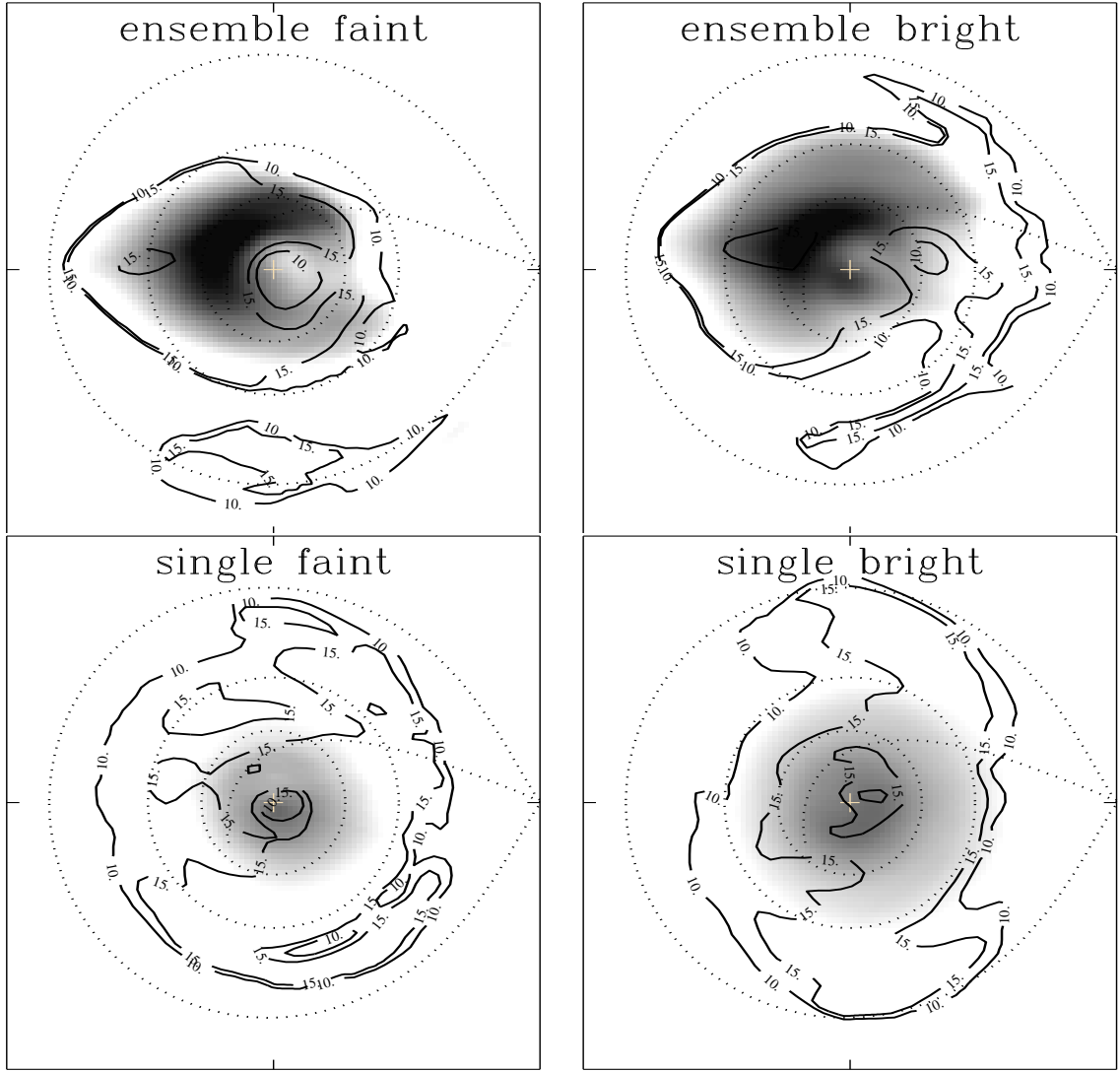}
\caption{ The statistical significance of the flickering maps.
   Contour lines for S/N=10 and 15 are overploted on the 
   corresponding `ensemble' and `single' maps of Fig.~\ref{fig6}. }
\label{fig12}
\end{figure}
%
\begin{figure}	
\includegraphics[bb=0cm 2cm 21cm 18cm,angle=-90,scale=0.6]{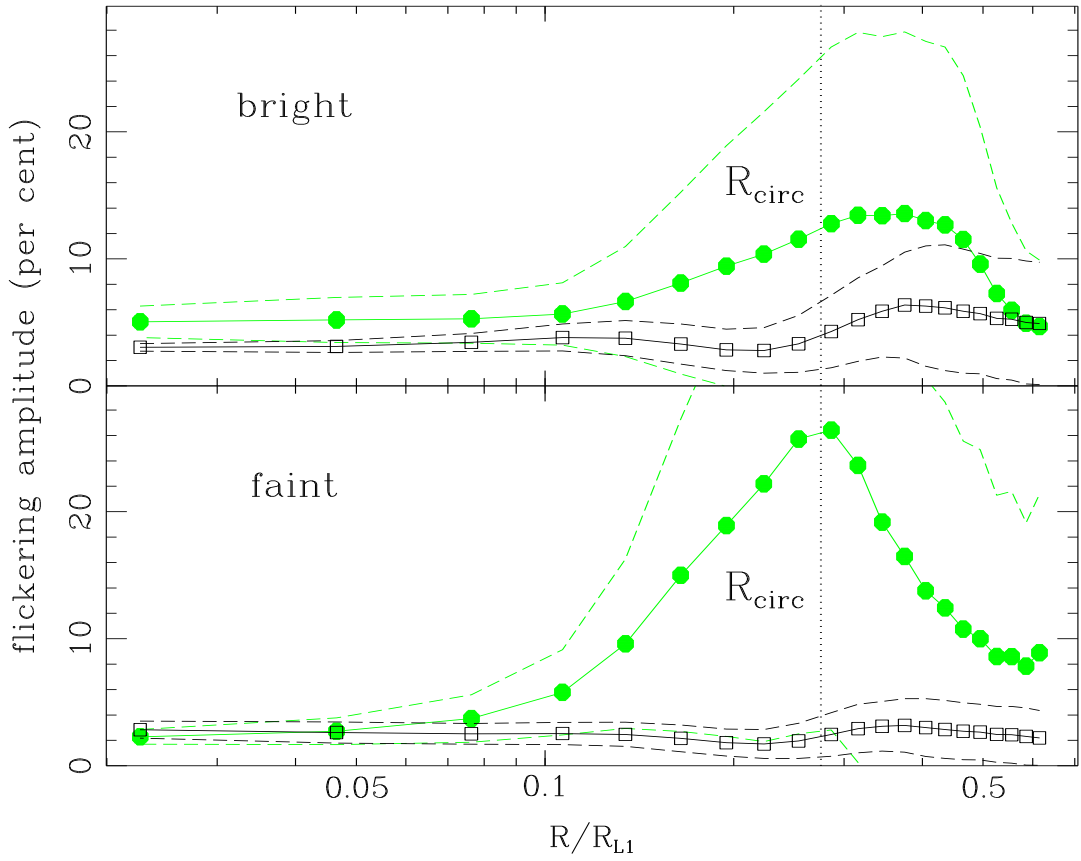}
\caption{ The relative amplitude of the flickering as a function of
   disk radius for the bright (top) and faint (bottom) states, from the
   'ensemble' (filled circles) and 'single' (open squares) maps. 
   The notation is the same as in Fig.~\ref{fig8}. }
\label{fig9}
\end{figure}
%
%
\begin{figure}	
\includegraphics[bb=1cm 2cm 21cm 18cm,angle=-90,scale=0.7]{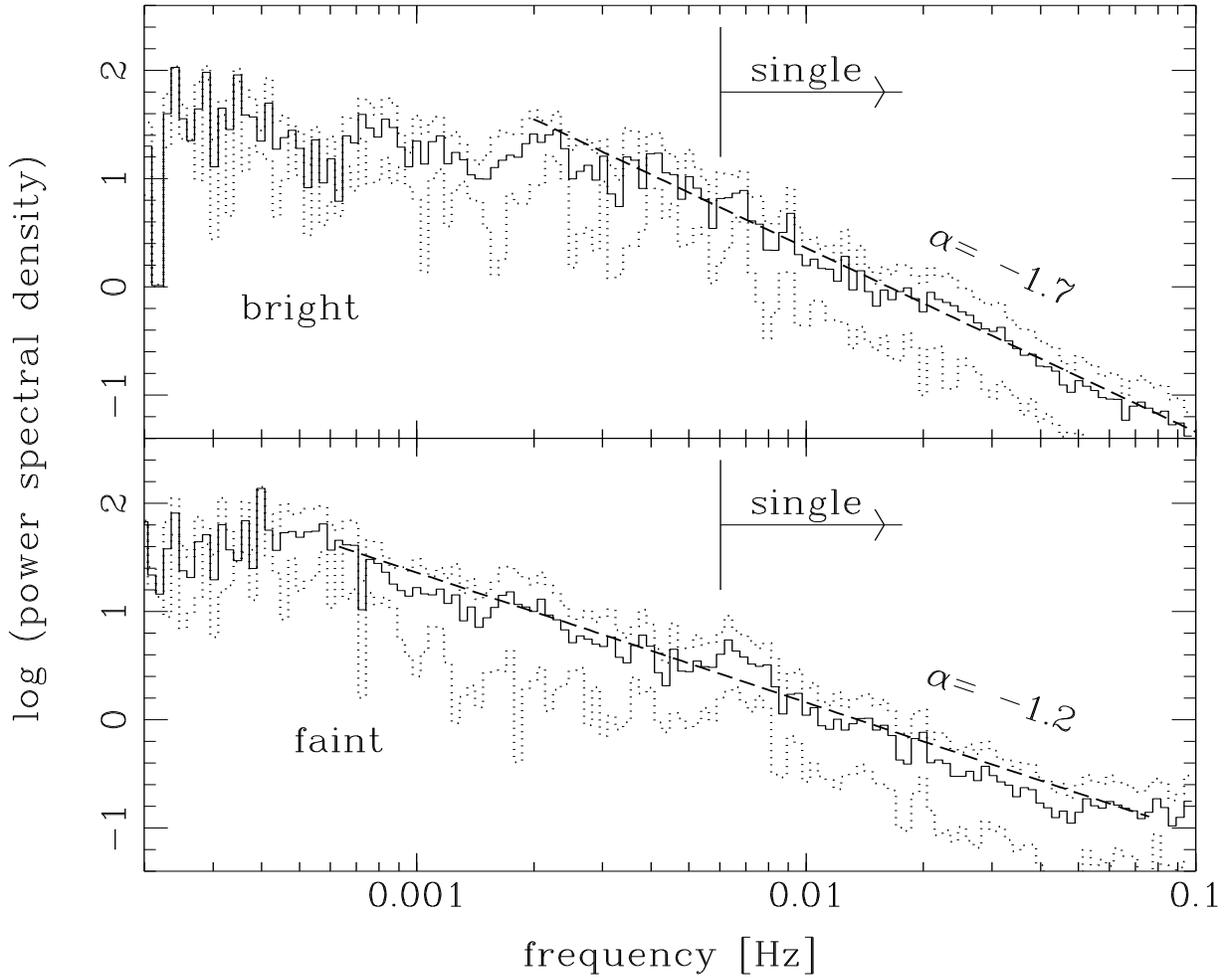}
\caption{ Average power spectrum density of the faint (bottom) and
   bright (top) states. The dotted lines show the 1-$\sigma$ limits on 
   the average power in each case. The best-fit power-law $P(f) \propto 
   f^{\alpha}$ is shown as a dashed line in each panel and the 
   corresponding slope is indicated. Vertical ticks mark the 
   low-frequency cut-off of the filtering process applied to derive 
   the `single' scatter curves. }
\label{fig10}
\end{figure}
%
%
\begin{figure}	
\includegraphics[bb=1cm 0cm 21cm 18cm,angle=-90,scale=0.6]{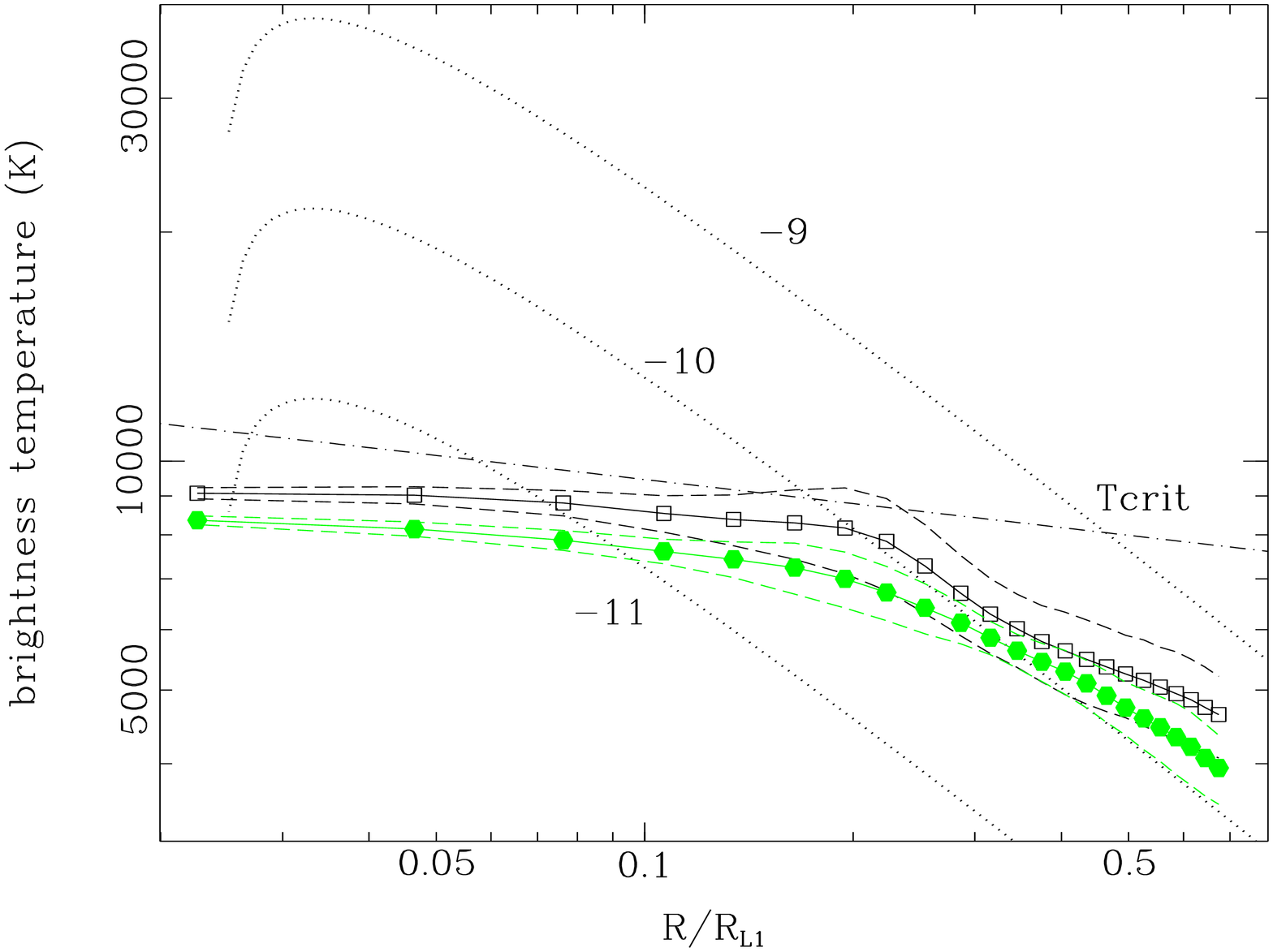}
\caption{ The brightness temperature radial distribution of the faint (filled
  circles) and bright (open squares) states, calculated assuming a distance 
  of 146\,pc to the system (Vrielmann et~al. 2002) and no reddening.  
  The dashed lines show the 1-$\sigma$ limits on the average temperatures.
  Steady-state disk models for mass accretion 
  rates of $\log$ \.{M}$= -9, -10$, and $-11 \;M_\odot\;$yr$^{-1}$ are 
  plotted as dotted lines for comparison. These models assume $M_1= 0.78\;
  M_\odot$ and $R_1= 0.0103\;R_\odot$ (Baptista et~al. 1998a). A dot-dashed 
  line marks the critical temperature above which the gas should remain in 
  a steady, high mass accretion regime according to the disc instability
  model (Warner 1995). }
\label{fig7}
\end{figure}
%


\begin{thebibliography}{99}

\bibitem {aug99} Augusteijn, T., et~al. 1992, A\&A, 265, 177
\bibitem {bap2001} Baptista, R. 2001, in Astrotomography: Indirect Imaging
		Methods in Observational Astronomy, eds.\ H. M. J. Boffin, D. Steeghs
		\& J. Cuypers (Berlin: Springer-Verlag), 307
\bibitem {bs93}  Baptista, R. \& Steiner, J. E. 1993, A\&A, 277, 331
\bibitem {bsh96} Baptista, R., Steiner, J. E. \& Horne, K. 1996, MNRAS, 282, 99
\bibitem {bap98} Baptista, R., Catal\'an, M. S., Horne, K. \& Zilli, D. 1998a, 
		MNRAS, 300, 233
\bibitem {14} Baptista, R., Horne, K., Wade, R., Hubeny, I., Long, K. \& Rutten,
		R. G. M. 1998b, MNRAS, 298, 1079
\bibitem {bhs00} Baptista, R., Harlaftis, E. T. \& Steeghs, D. 2000, MNRAS, 
		314, 727
\bibitem {bbh02} Baptista, R., Bortoletto, A. \& Harlaftis, E. T. 2002, MNRAS,
		335, 665
\bibitem {b03} Baptista, R., Borges, B. W., Bond, H. E., Jablonski, F., Steiner, 
		J. E. \& Grauer, A. D. 2003, MNRAS, 345, 889
\bibitem {bennie} Bennie, P. J., Hilditch, R. \& Horne, K. 1996, in Cataclysmic
		Variables and related objects, IAU Coll. 158, eds. A. Evans 
		\& J. Wood (Dordrecht: Kluwer), 33
\bibitem {bessell} Bessell, M. A. 1990, PASP, 102, 1181
\bibitem {bob97} Bobinger, A., Horne, K., Mantel, K. H. \& Wolf, S. 1997, A\&A,
		327, 1023
\bibitem {bruch1992} Bruch, A. 1992, A\&A, 266, 237
\bibitem {bruch1996} Bruch, A. 1996, A\&A, 312, 97
\bibitem {bruch2000} Bruch, A. 2000, A\&A, 359, 998
\bibitem {catalan} Catalan, M. S., Horne, K., Cheng, F.-H., Marsh, T. R. \& 
		Hubeny I. 1998, in Wild Stars in the Old West, ASP Conf.\ Series 137, 
		p. 426
\bibitem {b8} Cook, M. C. \& Brunt, C. C. 1983, MNRAS, 205, 465
\bibitem {25} Frank, J., King, A. R., \& Raine, D. J. 1992, Accretion Power in
			Astrophysics - 2nd edition (Cambridge: Cambridge University Press)
\bibitem {b32} G\"ansicke, B. T., Beuermann, K. \& de Martino, D. 1995, A\&A,
		303, 127
\bibitem {agn} Garcia, A., Sodr\'e, L., Jablonski, F. J. \& Terlevich, R. J.
		1999, MNRAS, 309, 803
\bibitem {ga} Geertsema, G. T. \& Achterberg, A. 1992, A\&A, 255, 427
\bibitem {godon} Godon, P. \& Sion, E. M. 2003, ApJ, 586, 427
\bibitem {graham} Graham, J. A. 1982, PASP, 94, 244
\bibitem {h2004} Harlaftis, E. T., Baptista, R., Morales-Rueda, L., Marsh, 
		T. R. \& Steeghs, D. 2004. A\&A, 417, 1063
\bibitem {h85} Horne, K. 1985, MNRAS, 213, 129
\bibitem {propeller} Horne, K. 1999, in Magnetic Cataclysmic Variables, 
		ASP Conf.\ Series Vol.\ 157, eds. K. Mukai \& C. Hellier (San 
		Francisco: ASP), 349
\bibitem {hs85} Horne, K. \& Stiening, R. F. 1985, MNRAS, 216, 933
\bibitem {hs} Herbst, W. \& Shevchenko, K. S. 1999, AJ, 118, 1043
\bibitem {kawa} Kawaguchi, T., Mineshige, S., Machida, M., Matsumoto, R. \&
		Shibata K. 2000, PASJ, 52, L1
\bibitem {lamla} Lamla, E. 1981, in Landolt-B\"{o}rnstein - Numerical Data 
		and Functional Relationships in Science and Technology, Vol.\,2, eds.\
		K. Schaifers \& H. H. Voigt (Berlin: Springer-Verlag)
\bibitem {lasota} Lasota, J.-P. 2001, New Astron. Reviews, 45, 449
\bibitem {lupton} Lupton, R. 1993, Statistics in Theory and Practice,
		(Princeton: Princeton University Press)
\bibitem {b53} Press, W. H., Flannery, B. P., Teukolsky, S. A. \& Vetterling, W.
		T. 1992, Numerical Recipes in C (Cambridge: Cambridge University Press)
\bibitem {44} Rutten, R. G. M., van Paradijs, J. \& Tinbergen, J. 1992, A\&A,
		260, 213
\bibitem {ss} Shakura, N. I. \& Sunyaev, R. A. 1973, A\&A, 24, 337
\bibitem {shu76} Shu, F. H. 1976, in Structure and Evolution of Close Binary
		Systems, IAU Symp. 73, eds.\ P. Eggleton, S. Mitton \& J. Whelan
		(Dordrecht: Reidel), 253
\bibitem {soko} Sokoloski, J. L., Bildsten, L. \& Ho W. C. G. 2001, MNRAS, 
		326, 553
\bibitem {sb} Stone, R. P. S. \& Baldwin, J. A. 1983, MNRAS, 204, 347
\bibitem {danny} Steeghs, D., O'Brien, K., Horne, K., Gomer, R. \& Oke, J. B.
		2001, MNRAS, 323, 484
\bibitem {vso} Vrielmann, S., Stiening, R. F. \& Offutt, W. 2002, MNRAS, 334, 608
\bibitem {w95} Warner, B. 1995, Cataclysmic Variable Stars (Cambridge:
		Cambridge University Press)
\bibitem {wn} Warner, B. \& Nather, R. E. 1971, MNRAS, 152, 219
\bibitem {wc} Warner, B. \& Cropper, M. 1983, MNRAS, 203, 909
\bibitem {wo} Warner, B. \& O'Donoghue D. 1987, MNRAS, 224, 733
\bibitem {ww} Welsh, W. F. \& Wood, J. H. 1995, in Flares and Flashes, eds.\
		J. Greiner, H. W. Duerbeck \& Roald E. Gershberg (Berlin:
		Springer-Verlag), 300
\bibitem{wip} Wood, J. H., Irwin, M. J. \& Pringle, J. E. 1985, MNRAS, 214, 475
\bibitem {wood86} Wood, J. H., Horne, K., Berriman, G., Wade, R., O'Donoghue, 
		D. \& Warner, B. 1986, MNRAS, 219, 629
\bibitem {wood89} Wood, J. H., Horne, K., Berriman, G. \& Wade, R. 1989, ApJ,
		341, 974

\end{thebibliography}
\end{document}